\documentclass[sn-mathphys,Numbered]{sn-jnl}

\usepackage{graphicx}%
\usepackage{multirow}%
\usepackage{amsmath,amssymb,amsfonts}%
\usepackage{amsthm}%
\usepackage{mathrsfs}%
\usepackage[title]{appendix}%
\usepackage{xcolor}%
\usepackage{textcomp}%
\usepackage{manyfoot}%
\usepackage{booktabs}%
\usepackage{algorithm}%
\usepackage{algorithmicx}%
\usepackage{algpseudocode}%
\usepackage{listings}%

\usepackage{subcaption}%
\usepackage{gensymb}

\theoremstyle{thmstyleone}%
\theoremstyle{thmstyletwo}%

\theoremstyle{thmstylethree}%

\usepackage[normalem]{ulem}
\usepackage{xspace}
\usepackage{comment}
\usepackage{orcidlink}

\newcommand{\pt}{\ensuremath{p_\text{T}}\xspace}

\begin{document}

\title[Article Title]{Application of Geometric Deep Learning for Tracking of Hyperons in a Straw Tube Detector}


\author*[1]{\fnm{Adeel} \sur{Akram} \orcidlink{0000-0003-0198-5852}}\email{adeel.akram@physics.uu.se}

\author[2]{\fnm{Xiangyang} \sur{Ju} \orcidlink{0000-0002-9745-1638}}\email{xju@lbl.gov}

\author*[1]{\fnm{Michael} \sur{Papenbrock} \orcidlink{0000-0003-0990-3145}}\email{michael.papenbrock@physics.uu.se}

\author[1]{\fnm{Jenny} \sur{Taylor} \orcidlink{0000-0002-9770-7488}\footnote{Current address GSI Helmholtzzentrum für Schwerionenforschung GmbH, Planckstr. 1, 64291 Darmstadt (Germany)}}\email{j.taylor@gsi.de}

\author[3]{\fnm{Tobias} \sur{Stockmanns} \orcidlink{0000-0002-6665-0095}}\email{t.stockmanns@fz-juelich.de}

\author*[1]{\fnm{Karin} \sur{Schönning} \orcidlink{0000-0002-3490-9584}}\email{karin.schonning@physics.uu.se}

\affil*[1]{\orgdiv{Department of Physics and Astronomy}, \orgname{Uppsala university}, \orgaddress{\street{Lägerhyddsvägen 1}, \city{Uppsala}, \postcode{752 37}, \state{}, \country{Sweden}}}

\affil[2]{\orgdiv{Physics Division}, \orgname{Lawrence Berkeley National Laboratory}, \orgaddress{\street{1 Cyclotron Road}, \city{Berkeley}, \postcode{CA 94720}, \state{California}, \country{United States}}}

\affil[3]{\orgdiv{Department of Physics}, \orgname{Forschungszentrum Jülich}, \orgaddress{\street{Wilhelm-Johnen-Straße}, \city{Jülich}, \postcode{52428}, \state{North Rhine-Westphalia}, \country{Germany}}}


\abstract{We present track reconstruction algorithms based on deep learning, tailored to overcome specific central challenges in the field of hadron physics. Two approaches are used: \textit{(i)} deep learning (DL) model known as fully-connected neural networks (FCNs), and \textit{(ii)} a geometric deep learning (GDL) model known as graph neural networks (GNNs). The models have been implemented to reconstruct signals in a non-Euclidean detector geometry of the future antiproton experiment PANDA. In particular, the GDL model shows promising results for cases where other, more conventional track-finders fall short: \textit{(i)} tracks from low-momentum particles that frequently occur in hadron physics experiments and \textit{(ii)} tracks from long-lived particles such as hyperons, hence originating far from the beam-target interaction point. Benchmark studies using Monte Carlo simulated data from PANDA yield an average technical reconstruction efficiency of 92.6\% for high-multiplicity muon events, and 97.1\% for the $\Lambda$ daughter particles in the reaction $\bar{p}p \to \bar{\Lambda}\Lambda \rightarrow \bar{p}\pi^+ p\pi^-$. Furthermore, the technical tracking efficiency is found to be larger than 70\% even for particles with transverse momenta \pt below 100 MeV/c. For the long-lived $\Lambda$ hyperons, the track reconstruction efficiency is fairly independent of the distance between the beam-target interaction point and the $\Lambda$ decay vertex. This underlines the potential of machine-learning-based tracking, also for experiments at low- and intermediate-beam energies.}




\maketitle

\section{Introduction}\label{intro}

The goal of \textit{hadron physics} is to understand how the strong interaction binds massless gluons and almost massless quarks into the massive hadrons that form our visible universe. The relevant interactions occur within the \textit{confinement domain}, which corresponds to an energy range of $\approx$100 MeV up to a few tens of GeV. Particles produced in this energy region typically give rise to complicated detector signatures, including high-curvature tracks from low-momentum particles. Furthermore, the curvature of these tracks changes along the trajectory due to energy loss and other interactions with the material, hence these trajectories cannot be described by simple geometries. Another challenge is the similarity between signatures of interest and background -- a feature particularly prevalent at the low and intermediate energies characterising hadron physics. Constructing software algorithms that can handle these features is therefore challenging -- but necessary to fully exploit the new and the next generations of hadron physics facilities. A prominent example of the latter is the future PANDA experiment\cite{PANDA:2009yku}, where a beam of stored antiprotons and a large-acceptance, multi-purpose detector will open new avenues in exploring the strong interaction. 

\textit{Hyperons} are similar to the protons but also contain at least one heavy and unstable strange quark. Due to their weak, self-analysing decays, hyperons provide a precise diagnostic tool to test CP symmetry \cite{BESIII:2018cnd, BESIII:2021ypr} and electromagnetic structure \cite{BESIII:2019nep, Schonning:2023mge}. Hyperon spectroscopy provides crucial information about how composite systems emerge from strongly interacting quarks and gluons \cite{Crede:2013kia, Thiel:2022xtb}, and interactions between hyperons, antihyperons and nuclei provide important pieces to the hyperon puzzle of neutron stars \cite{Tolos:2020aln}.  In PANDA, the antiproton-proton annihilations will enable the production of all known and predicted single-, double- and triple-strange hyperons ($Y$) in two body reactions $\bar{p}p \to \bar{Y}Y$. This provides a clean, particle-antiparticle symmetric final state that is straightforward to parametrise -- a significant advantage for spectroscopic partial wave analyses \cite{PANDA:2020hmi}, investigations of interaction dynamics \cite{Singh:2016hoh, PANDA:2020zwv} and CP symmetry tests \cite{Schonning:2023mge}. Simulation studies have demonstrated that PANDA will be a \textit{strangeness factory} already in its first operation phase \cite{PANDA:2021ozp}. 

Identifying hyperons has its unique challenges. Ground-state hyperons decay weakly, on a time-scale of $10^{-10}$ s. This implies that relativistic particles travel a distance of a few centimetres or even meters before decaying. Therefore, they will leave a track in the detector that starts a finite distance away from the beam-target interaction point (IP), \textit{i.e.} a \textit{displaced vertex}. As an example, the flight distance of the $\Lambda$ hyperon is $c\tau = 7.89 \textrm{ cm}$. Heavier hyperons, such as the multi-strange $\Xi^0$, $\Xi^-$, $\Xi^*$, $\Omega^-$ and $\Omega^*$, decay by considerable fractions into states containing a $\Lambda$ hyperon. Prominent decay channels of charm baryons, such as the $\Lambda_c^+$, also contain the ground-state strange $\Lambda$ or $\Sigma$ hyperons. Hence, successful hyperon analyses rely on the ability to reconstruct tracks from displaced vertices and at the same time, to properly handle highly curved or even spiralling tracks from low-momentum particles \cite{PANDA:2023ljx}. However, most generic tracking algorithms are tailored to primary tracks, since the assumption that a track originates from the IP is a powerful constraint that reduces combinatorics and the background, as well as improves the momentum resolution. This is also the case for the standard PANDA track finder, used in most non-hyperon analyses up to now (see \textit{e.g.} Refs.~\cite{PANDA:2016fbp, PANDA:2020jkm, PANDA:2018zjt}). In recent years, several efforts have been made to develop generic algorithms that can handle secondary tracks from displaced vertices \cite{Andersson:2020grb, Taylor:2024rqt, Alicke:2023lnt}. These are all based on classical approaches such as the Hough transformation, recursive annealing filters and Apollonius triplets. In particular, the algorithms explored by Alicke in Ref. \cite{Alicke:2023lnt} improve the reconstruction efficiency for secondary tracks up to 79\%, to be compared with 45\% for the standard primary track finder. Nevertheless, there is room for improvement and it is notable that Refs. \cite{Alicke:2023lnt, Andersson:2020grb, Taylor:2024rqt} do not explicitly address low-momentum particles. 


Machine-learning techniques are gaining importance in particle tracking, sparked by the \textit{Tracking Machine Learning Challenge (TrackML)} within the high-energy community \cite{Amrouche:2019wmx, Amrouche:2021nbs}. In particular, Graph Neural Networks (GNN) have been found suitable for particle tracking in detectors with non-Euclidean geometries such as ATLAS \cite{ExaTrkX:2021abe, Caillou:2022hly}, WASA-FRS at FAIR \cite{Ekawa:2023nzz}, BESIII \cite{Jia:2024rbx} and Belle II \cite{Reuter:2024kja}. Since the central part of the PANDA detector has a non-Euclidean geometry, GNNs are natural candidates for facing the challenge of its track reconstruction. It is crucial to investigate how GNN-based solutions tackle the specific challenges of a low-to-intermediate energy experiment such as PANDA, particularly low-momentum particles with displaced vertices. In this paper, we present a detailed ML-based track reconstruction solution for a non-Euclidean PANDA detector system. 

The paper is organized as follows. Section \ref{sec:panda} briefly introduces the PANDA experiment at FAIR with an emphasis on the Straw Tube Tracker, Section \ref{sec:trackml} gives an overview of track reconstruction using machine learning, including an account of contemporary related work. Section \ref{sec:methodology} presents a detailed description of our methodology, from machine learning architecture designs to the performance evaluation metrics. Section \ref{sec:results} gives the final results for three different cases: \textit{(i)} muon reconstruction using conventional deep learning, \textit{(ii)} muon reconstruction using geometric deep learning, and \textit{(iii)} hyperon reconstruction using geometric deep learning. Finally, in Section \ref{sec:conclusion} we present the conclusions.

\section{The PANDA Experiment at FAIR} \label{sec:panda}

\begin{figure}[!htb]
  \centering
  \includegraphics[width=\linewidth]{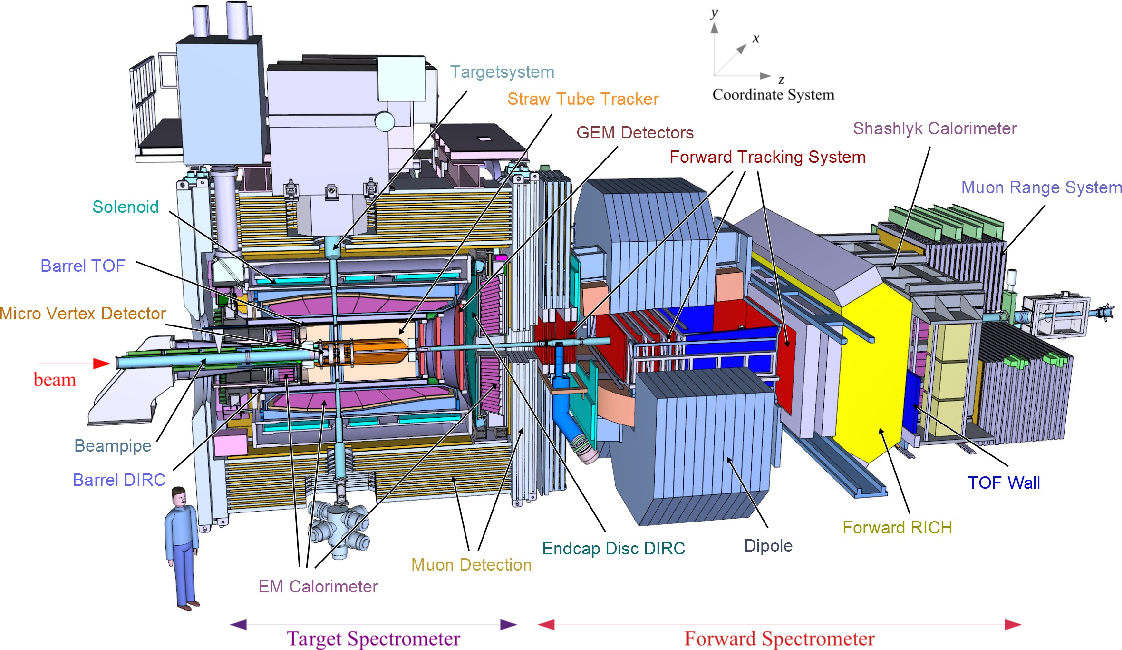}
  \caption{Schematic of the PANDA Experiment}
  \label{fig:panda}
\end{figure}

The PANDA (anti-Proton ANnihilation at DArmstadt) experiment is currently under construction at FAIR (Facility for Anti-proton and Ion Research) \cite{Gutbrod:54062, Gutbrod:54068}. The antiproton beam from the High Energy Storage Ring (HESR) will impinge on a fixed, internal hydrogen cluster jet target, a hydrogen or deuterium pellet target (for $\bar{p}p$ or $\bar{p}n$ reactions) or foils (for $\bar{p}A$ interactions). The interaction rate will be 1-20 MHz. A schematic layout of the planned detector is shown in Fig. \ref{fig:panda}. 

\noindent The detector will consist of two parts: a target spectrometer (TS) employing a solenoid magnet for the detection of particles emitted centrally, and a Forward Spectrometer (FS) with a dipole magnet for forward-going particles. Together, the two spectrometers cover almost the full 4$\pi$ solid angle. In this work, we focus on the TS that comprises a Micro Vertex Detector (MVD) surrounding the IP, followed by a Straw Tube Tracker (STT) for tracking. A Barrel Detector of Internally Reflected Cherenkov light (DIRC) and a Barrel Time-of-Flight (ToF) provide particle identification and an electromagnetic calorimeter will measure the energies of charged and neutral particles. The solenoid magnet will surround the EMC and its iron yoke will act as an absorber. Most particles that traverse the full iron yoke will be muons, and they will be tagged in a dedicated muon system. The full PANDA detector is described in detail in Refs.~\cite{PANDA:2009yku, PANDA:2021ozp}.

\begin{figure}[!htb]
  \centering
  \includegraphics[width=0.8\linewidth]{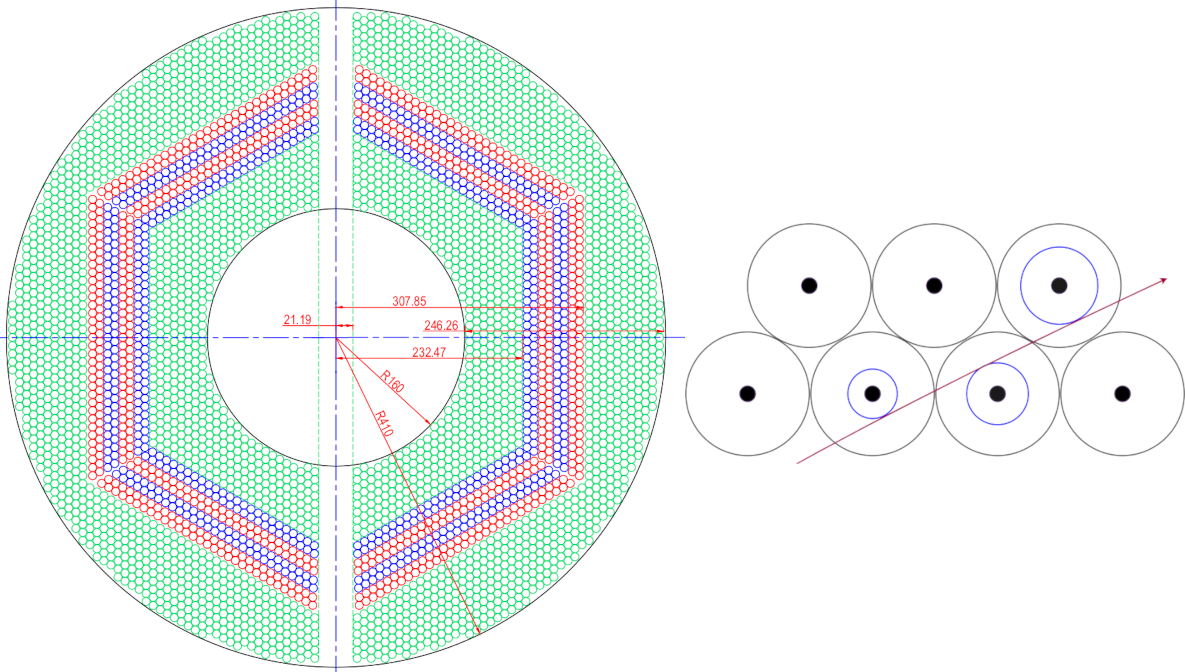}
  \caption{Left: a cross-sectional view of the STT detector. Right: zoom-in view of straw tubes in black circles, the isochrone radius in blue circles, and a track in red.}
  \label{fig:stt}
\end{figure}

In this work, we focus on data collected by the STT, which will consist of 4224 single-channel straw tubes, distributed in 27 layers and six sectors in a hexagonal shape, as shown in Fig.~\ref{fig:stt} (left). The green-marked tubes (15 - 19 layers) will be arranged parallel to the beam axis to measure the hit position, whereas blue- and red-marked tubes (8 layers) are tilted or \textit{skewed} with respect to the beam axis by a $\pm 2.9^{\degree}$ polar angle. The skewed straws enable reconstruction of the $z-$component of the tracks~\cite{Andersson:2020grb}. The STT detector will cover polar angles from $\theta = 22^{\degree}$ to $\theta = 140^{\degree}$. For a solenoid field of 2 T, particles produced at the IP with a transverse momentum \pt of at least 50 MeV will reach the innermost layer while a minimum \pt of 100 MeV is required to traverse through the full STT.

\noindent When a charged particle traverses the STT detector, it ionises the gas inside the tube, releasing electrons. The electrons then drift toward the centre of the tube, ionising more gas molecules on the way. All free electrons will be collected at an anode wire in the centre, resulting in a signal pulse referred to as a \textit{hit}. The $xy$ position of the hit is the position of the anode wire. The distance of the closest approach from the particle trajectory to the anode wire is the \textit{isochrone radius}. Our analysis uses the $xy$ position from the straight straws and the isochrone radii as input data.

\subsection{PandaRoot Analysis Framework}

We used the PandaRoot analysis framework \cite{Spataro_2011} to produce the simulated data for the analysis. PandaRoot offers tools for event simulation, beginning with the production of Monte Carlo events and continuing with the propagation of particles through detector material, digitisation of signals, reconstruction and calibration, and physics analysis. PandaRoot, a detector-specific framework, is derived from the general-purpose FairRoot framework \cite{Al-Turany:2012zfk}, which in turn is based on the ROOT framework \cite{Brun:1997pa}. FairRoot constitutes a base for other detector-specific frameworks within the FAIR software ecosystem and provides a wide range of basic classes that facilitate the customising of each detector configuration. Furthermore, it provides an event display, database management, an input-output manager, a run manager, and the Virtual Monte Carlo (VMC) interface. The latter enables the selection of several simulation engines. In addition, it uses the task system of ROOT to combine and exchange different algorithms into a simulation chain.

\section{Track Reconstruction with ML}\label{sec:trackml}

Track reconstruction, a process that labels hits to reconstruct particle trajectories, is essential for almost all physics analyses. Various algorithms have been developed, most tailored to the experiment and the particles at focus. Factors such as particle multiplicity, point of origin, and momentum are crucial when choosing an algorithm. For instance, in some high-energy physics (HEP), particles produced in a beam-target interaction have large transverse momenta \pt, resulting in mostly straight particle trajectories. In contrast, hadron physics experiments often produce particles with \pt as low as 100 MeV/c, leading to highly curved particle trajectories that may intersect multiple other particle trajectories in the detector. This factor alone makes track reconstruction a demanding task. Another factor is that long-lived particles such as hyperons decay centimetres or even meters from the beam-target interaction point. On the other hand, the track multiplicity is generally lower in hadron physics compared to HEP; it is unusual that one $\bar{p}p$ annihilation event within the energy region of PANDA gives rise to more than ten particles. 

At the heart of every algorithm is the pattern recognition algorithm. Most \textit{classical} algorithms~\cite{PANDA:2016fbp, PANDA:2020jkm, PANDA:2018zjt} are combinatorial; they recursively try different hit combinations to find particles, which makes these algorithms computationally expensive. In this work, we explore an ML-based solution for track reconstruction to address not only the computational challenges but also the aforementioned challenges of track reconstruction in hadron physics experiments.


\subsection{Related Work}\label{sec:relatedwork}

Pattern recognition using neural networks has seen significant advancements in recent years. Within the HEP.TrkX project \cite{heptrkx}, novel deep learning techniques were developed for track reconstruction to address the challenges of High-Luminosity LHC (HL-LHC). These solutions considered image-based techniques, such as image segmentation and image captioning, recurrent neural networks (RNNs) and convolutional neural networks (CNNs), applied to a pseudo-data from simulations of a planar detector geometry \cite{Farrell2017Aug}. However, these methods do not scale with realistic detectors of irregular geometries and data sparsity. Using the space-point representation of tracking data from a generic barrel detector, RNNs and GNNs were used for track reconstruction with great success \cite{Tsaris2017Aug, Farrell2018Oct}. Building upon these developments, the Exa.TrkX project \cite{exatrkx} which is the successor of HEP.TrkX, demonstrated the potential of GNNs to broader particle track and shower reconstruction \cite{Ju2020Mar}, track-seeding and labelling \cite{Choma2020Jun} including full-detector analyses \cite{Biscarat2021Mar}. Additional applications of GNNs in particle physics can be found in Refs. \cite{Pata2021May, Ju2021Oct, DeZoort2021Nov}. Almost all these applications use the TrackML data simulated with a generic detector geometry. For application to the more realistic detector is reported in Refs. \cite{Caillou:2022hly, Caillou:2024smf}. Beyond HL-LHC, the GNNs have been used in several other realistic detectors such as GEM detectors \cite{Baranov2019Oct}, straw-tube detectors \cite{Esmail2022, Akram2022Aug, Akram:2023eci}, drift chambers \cite{Ekawa2023May, Jia2024May, Reuter2024Nov}, and LArTPCs \cite{Hewes2021Mar, Aurisano2024Mar}. To keep track of the applications of GNNs in nuclear and particle physics, we refer readers to the \textit{HEP ML Living Review} \cite{hepmllivingreview}.

In the PANDA experiment, GNN models have been utilized for edge classification in track reconstruction within the Straw Tube Tracker (STT), building upon the methodology outlined in Ref. \cite{Ju2021Oct}, with preliminary results presented in Ref. \cite{Akram2022Aug}. This paper takes a step further by exploring the application of various neural network architectures to data involving muons and hyperons with detailed studies presented in Ref. \cite{Akram:2023eci}.

\section{Methodology}\label{sec:methodology}

In this work, we aim to perform pattern recognition using concepts of \textit{nodes}, \textit{edges}, \textit{graphs}, and deep neural networks. It is natural to consider particle trajectories in a detector as graphs, where the graph nodes represent the detector hits and graph edges represent the possibility of two detector hits coming from the same particle. An edge is labelled as true if the two linked hits are from the same particle and false otherwise. The core idea is to build a graph from the detector hits that includes all true edges and as few as possible fake edges. Then, the graph can be classified by FCNs or GNNs. One can perform classification on graphs in three different ways: \textit{(a)} node classification where the hits are classified as either signal or noise, \textit{(b)} edge classification where a link between two hits is classified as either true or false, and \textit{(c)} graph classification where a full event is classified either signal or noise. For track reconstruction, an edge classification is a suitable option where all edges in a graph are labelled with \textit{edge scores}. The labelled graph is then passed to a clustering algorithm to group hits as track candidates.

First, we consider two representations of the data from the PANDA STT detector: Euclidean and non-Euclidean. They differ in whether a classification model is itself geometrical or non-geometrical. Second, we use two different reactions to produce events: muon pairs and hadrons from a $\bar{p}p \rightarrow \bar{\Lambda}\Lambda \rightarrow \bar{p}\pi^{+}p\pi^{-}$ reaction at beam momentum of 1.642 GeV/c. The latter corresponds to an excess energy of  $\sim 73$ MeV with respect to the $\bar{\Lambda} \Lambda$ production threshold and has been studied in detail before in the context of PANDA with ideal tracking \cite{PANDA:2020zwv}. Our strategy is to first compare both representations using hit data produced by the muons, and then use the best-performing approach to reconstruct hadrons from $\bar{p}p \rightarrow \bar{\Lambda}\Lambda \rightarrow \bar{p}\pi^{+}p\pi^{-}$ reaction. Here, tracks from the final state particles originate from the $\Lambda$ and $\bar{\Lambda}$ decay vertices, typically located several centimetres away from the beam-target interaction point. Hence, this reaction provides a benchmark for evaluating the performance of machine learning algorithms in reconstructing tracks from displaced decay vertices. 

The following sections detail the three major steps in our study: data generation and acquisition, the application of deep learning, and the evaluation of reconstructed trajectories. 

\subsection{Data Generation}

The muon data sample consists of five $\mu^\pm$ pairs in the momentum range from 100 MeV/c to 1.5 GeV/c, where the muons are produced with a particle gun, isotropically distributed in the STT acceptance. The $\bar{p}p \rightarrow \bar{\Lambda}\Lambda \rightarrow \bar{p}\pi^{+}p\pi^{-}$ reaction is simulated with the generator EvtGen \cite{Ryd:2005zz}. GEANT4 handles the particle transport through the detector material \cite{GEANT4:2002zbu}. For both muons and hyperons, $10^5$ events are generated for training which is further distributed into $90\%$ for training, $5\%$ for validation and $5\%$ for testing, respectively. For inference, a separate prediction sample containing $2 \cdot 10^3$ events for muons and $3 \cdot 10^3$ events for hyperons is used to avoid any bias.

Finally, we preprocess the events by creating a hit feature vector for each hit using their positions ($r, \phi, z$) and respective isochrone radii and defining true edges between hits with the Monte Carlo truth information. 

\subsection{Deep Learning Pipeline}\label{sec:pipeline}

\begin{figure}[!ht]
    \centering
    \includegraphics[width=\linewidth]{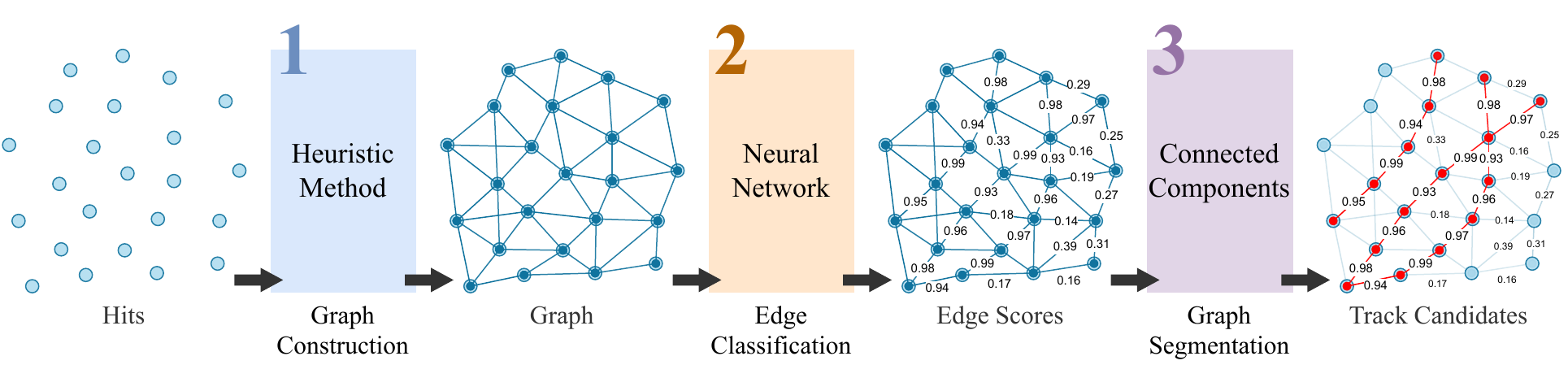}
    \caption{Sketch of a deep learning pipeline, image adapted from the Exa.TrkX~\cite{Ju2021Oct}.}
    \label{fig:pipeline}
\end{figure}

The deep learning pipeline contains several stages: \textit{(i)} Graph Construction, \textit{(ii)} Edge Classification, and \textit{(iii)} Graph Segmentation as shown in Fig.~\ref{fig:pipeline}.

\noindent In the \textbf{graph construction stage}, we use a \textit{layerwise} heuristic method to build edges by connecting nodes in adjacent layers, starting from the innermost layer to the outermost layer of the STT. If a node is missing in one layer, the edge is created with the next available layer. We also restrict graph construction in the adjacent sectors of the STT. Since the detector occupancy from hyperons is much smaller than the muons, we remove this constraint for hyperons. This stage produces an edge list where each edge is labelled as true or false depending on whether it belongs to a particle or not. Furthermore, all edges resulting from the noise, if any, are labelled as false.

In the \textbf{edge classification stage}, we train a deep learning model to classify edges as produced in the previous stage. This stage has two modes: Euclidean and non-Euclidean. In the Euclidean mode, each edge is fed separately to an FCN for classification. The relational information between hits beyond one-hop connections is inherently absent in this mode.  In the non-Euclidean mode, the full event presented as a graph, is fed to a GNN for classification. The idea is that the GNN can better capture the topological features of data. 

Our FCN model has six fully connected layers with hidden dimensions of \texttt{[128, 128, 1024, 1024, 128, 1]}. We apply the \texttt{relu()} activation function in hidden layers and the \texttt{sigmoid()} activation function in the final layer for binary classification. In addition, layer norm is applied to the final layer.

Similarly, the GNN model is the Interaction Graph Neural Network (IGNN)~\cite{Battaglia2016} formulated under a message-passing framework \cite{Battaglia2018}. The IGNN consists of three modules: \textit{(i)} encoder module, \textit{(ii)} graph module, and  \textit{(iii)} decoder or output module. The encoder module consists of an edge network and a node network, and its task is to encode input node features to a vector of hidden features and to create edge features from neighbouring nodes. In the graph module, aggregated neighbouring edge features are passed to the node network, and the neighbouring node features are passed to the edge network. This is a message-passing step where information is exchanged between nodes and edges. This step is repeated eight times. The final output is then passed to the output module that performs binary classification using the binary-cross-entropy loss function. As a result, each edge is assigned an edge score. The architecture, \textit{i.e.} nodes and layers of each module, is chosen exactly as used in Ref. \cite{Ju2021Oct}. A schematic diagram of IGNN used in this work is shown in Fig.~\ref{fig:ignn}:

\begin{figure}[!ht]
    \centering
    \includegraphics[width=0.9\linewidth]{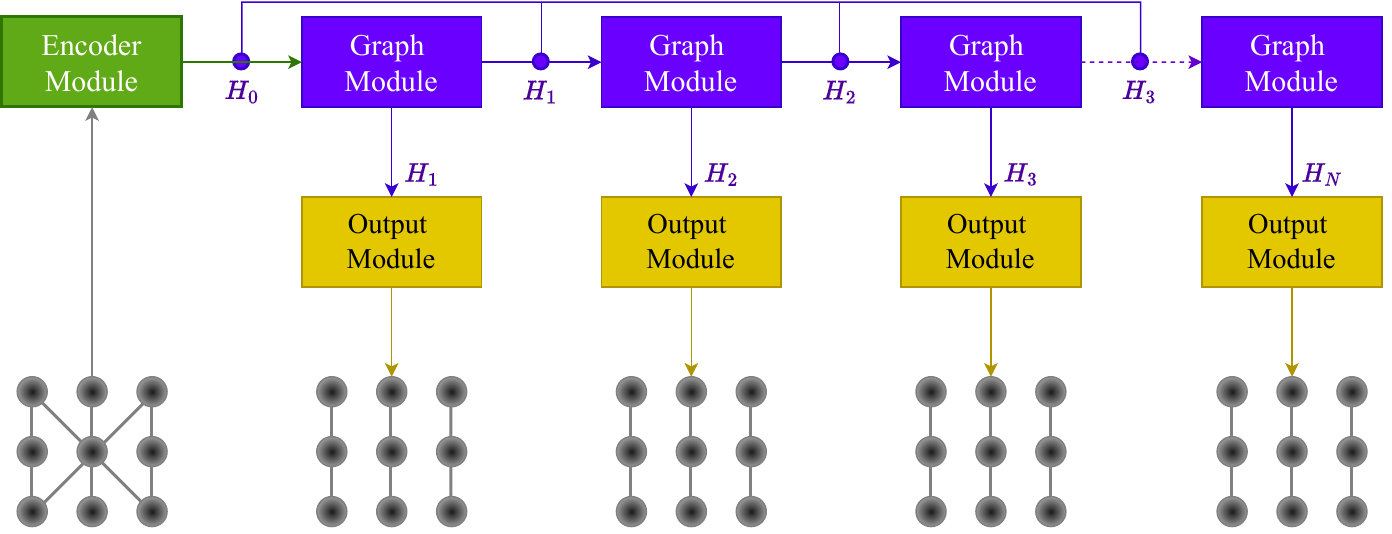}
    \caption{Schematic of an IGNN showing encoder, graph and output modules with $N=8$ message-passing steps. }
    \label{fig:ignn}
\end{figure}

\noindent Networks inside IGNN modules are FCNs. Each network has a three-layer architecture with nodes of \texttt{[128, 128, 128]} with a ReLU activation function on all layers. The following hyperparameters are used during training:

\begin{itemize}
    \item Binary Cross-Entropy (BCE) loss
    \item AmsGrad Optimizer: $\alpha=0.001, \beta_{1}=0.9, \beta_{2}=0.999$, weight\_decay=0.01
    \item Learning rate: $\alpha=0.001$
\end{itemize}

\noindent We set the batch size to 128 for the FCN and 1 for the IGNN. The networks are trained for $50$ epochs as the generalization gap between the training and the validation errors does not change significantly. Finally, each edge in the graph is labelled with a probability score later called the edge score.


In the \textbf{graph segmentation stage}, we use the density-based spatial clustering of applications with noise (DBSCAN) algorithm~\cite{Ester1996} to find connected components of the labelled graph. We use the graph score as the distance metric between two nodes. The distance metric ($\epsilon_{db}$) defines the maximum distance between two nodes to cluster them together. The value of $\epsilon_{db}$ is scanned to find an optimal value where the efficiency of graph segmentation is high. 

\subsection{Track Evaluation}\label{sec:track_eval}

Track evaluation ensures that the reconstructed tracks accurately represent true particle trajectories. One method to assess the quality of the track reconstruction algorithm is by calculating the overall tracking efficiency, referred to here as \textit{physics efficiency} and \textit{track purity}. Physics efficiency indicates how effectively the tracking algorithm can identify all particle tracks from the detector signals, and depends both on the performance of the algorithm and the efficiency of the detector. Track purity refers to the algorithm's ability to distinguish true particle tracks from wrongly reconstructed tracks, or types of other backgrounds. To evaluate the performance of the tracking algorithm itself, independently of the detector efficiency, a conditional tracking efficiency is defined, \textit{i.e.} the \textit{technical efficiency}. When this is evaluated, a minimum number of hits is required below which the algorithm cannot be expected to reconstruct a track. The tracking metrics will be defined using an evaluation scheme that closely aligns with the ATLAS community \cite{ATLASItk}:

\begin{itemize}
    \item \textbf{$N_{\text{particles}} (\text{selected})$} is the number of generated particles in the detector acceptance, which will be referred to as \textit{particles}.
    \item \textbf{$N_{\text{particles}} (\text{selected}, \text{matched})$} is the number of particles matched to at least one reconstructed track.
    \item \textbf{$N_{\text{particles}} (\text{selected}, \text{reconstructable})$} is the number of generated particles that leave at least seven or more hits ($N_t$) in the detector, they will be referred to as the \textit{reconstructable particles}.
    \item \textbf{$N_{\text{particles}} (\text{selected}, \text{reconstructable}, \text{matched})$} is the number of reconstructable particles that are matched to at least one reconstructed track.
    \item \textbf{$N_{\text{tracks}}(\text{selected})$} is the number of reconstructed tracks containing at least five or more hits ($N_r$), which will be referred to as \textit{reconstructed tracks}.
    \item \textbf{$N_{\text{tracks}}(\text{selected}, \text{matched})$} is the number of reconstructed tracks that are matched to a particle.
\end{itemize}

\noindent A particle is considered matched to a reconstructed track if more than \textit{(i)} 50\% of the hits in the reconstructed track belong to the same true particle, \textit{(ii)} 50\% of the hits in the matched true particle are found in the reconstructed tracks. This is known as \textit{two-way matching} scheme. Furthermore, the reconstructable particles are the selected particles that also have at least seven hits in the detector before performing the track reconstruction.

The physics efficiency ($\epsilon_{\textrm{phys}}$) is the fraction of particles that match at least one reconstructed track:

\begin{align}
    \epsilon_{\textrm{phys}} &= \frac{N_{particles} (\textrm{selected, matched})}{N_{particles} (\textrm{selected})}
\end{align}

\noindent The technical efficiency ($\epsilon_{\textrm{tech}}$) is the fraction of reconstructable particles that match at least one reconstructed track:

\begin{align}
    \epsilon_{\textrm{tech}} &= \frac{N_{particles} (\textrm{selected, reconstructable, matched})}{N_{particles} (\textrm{selected, reconstructable})}
\end{align}

\noindent Finally, the track purity ($\rho$) is defined as the fraction of reconstructed tracks that match a selected particle:

\begin{align}
    \rho &= \frac{N_{\text{tracks}} (\textrm{selected, matched})}{N_{\text{tracks}} (\textrm{selected})}
\end{align}

\noindent In addition, the \textit{fake rate} ($\equiv 1 - \rho$) or \textit{ghost rate} is defined as the fraction of reconstructed tracks not matching any particle tracks. In contrast, the \textit{clone rate} is the rate at which a particle is matched to more than one reconstructed track.


In addition to the requirement that the reconstructable track has at least 7 STT hits (\textit{i.e.} $N_t \ge 7$), reconstructed tracks require at least 5 STT hits (\textit{i.e.} $N_r \ge 5$) and a matching fraction (MF) greater than 50\%.

\section{Results}\label{sec:results}

For a fine-grained understanding of the performance, we investigate how the track efficiencies depend on variables such as transverse momentum \pt, and the radial distance $d_0$ between the beam-target interaction point and the decay vertex: 

\begin{equation}\label{eq:d0}
   \quad d_0 = \sqrt{v_x^{2} + v_y^2}.
\end{equation}

\noindent where $v_x$ and $v_y$ denote the positions of the decay vertex of particles along the X and Y axis, respectively.

\subsection{Muon Reconstruction with GDL}\label{sec:muon_gdl}

To reconstruct muons, two different approaches are adopted: the Euclidean, using FCN and the non-Euclidean, using IGNN. We investigate the performance of edge labelling and graph segmentation stages leading to the evaluation of both approaches.

\begin{figure}[!htb]
    \centering
    \subfloat{\includegraphics[width=0.49\linewidth]{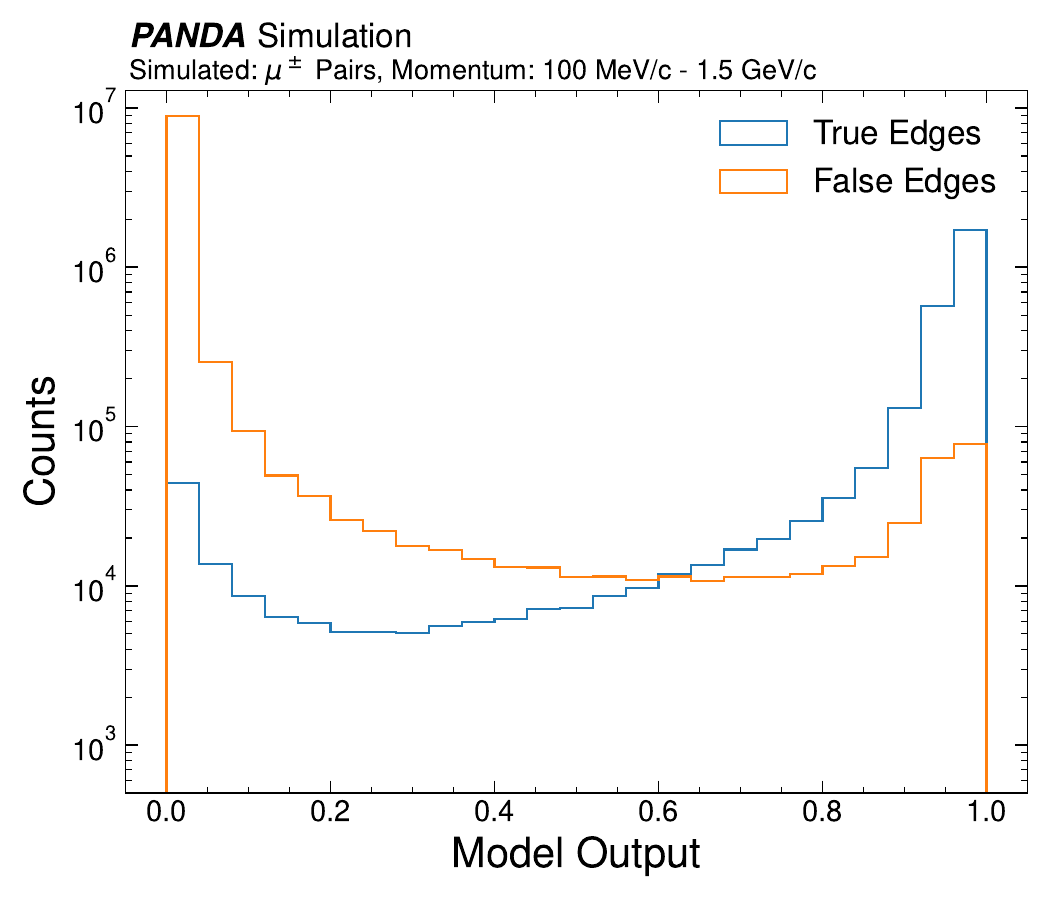}}
    \hfill
    \subfloat{\includegraphics[width=0.49\linewidth]{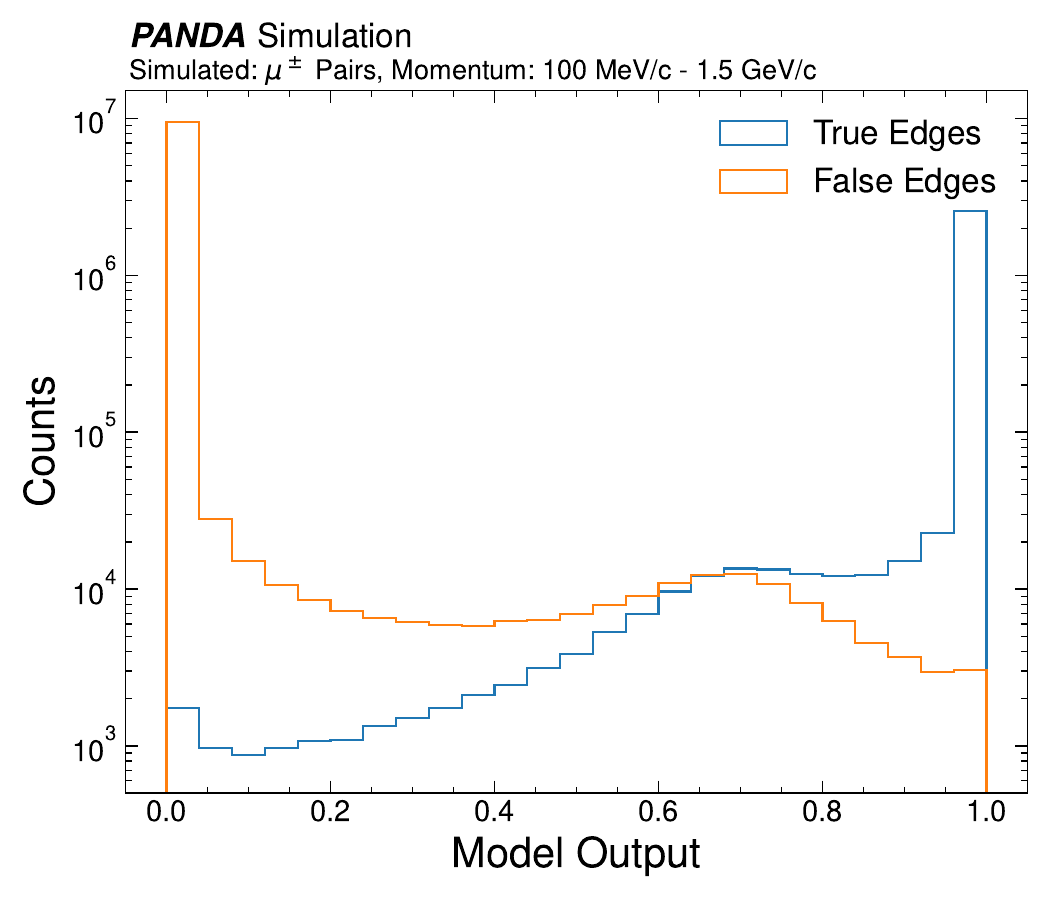}}
    \caption{The output of a classification model: FCN (left) and IGNN (right).}
    \label{fig:muon_model_ouputs}
\end{figure}

To examine the edge labelling different evaluation metrics are used. The model output gives the classification probabilities, referred to as edge scores, for each edge in the graph. For edge predictions, an optimal threshold on the edge score is required. Fig. \ref{fig:muon_model_ouputs} shows the model outputs of FCN and IGNN where edge scores of true (blue) and false (orange) edges are shown without applying a threshold on the model outputs.

\begin{figure}[!htb]
    \centering
    \subfloat{\includegraphics[width=0.49\linewidth]{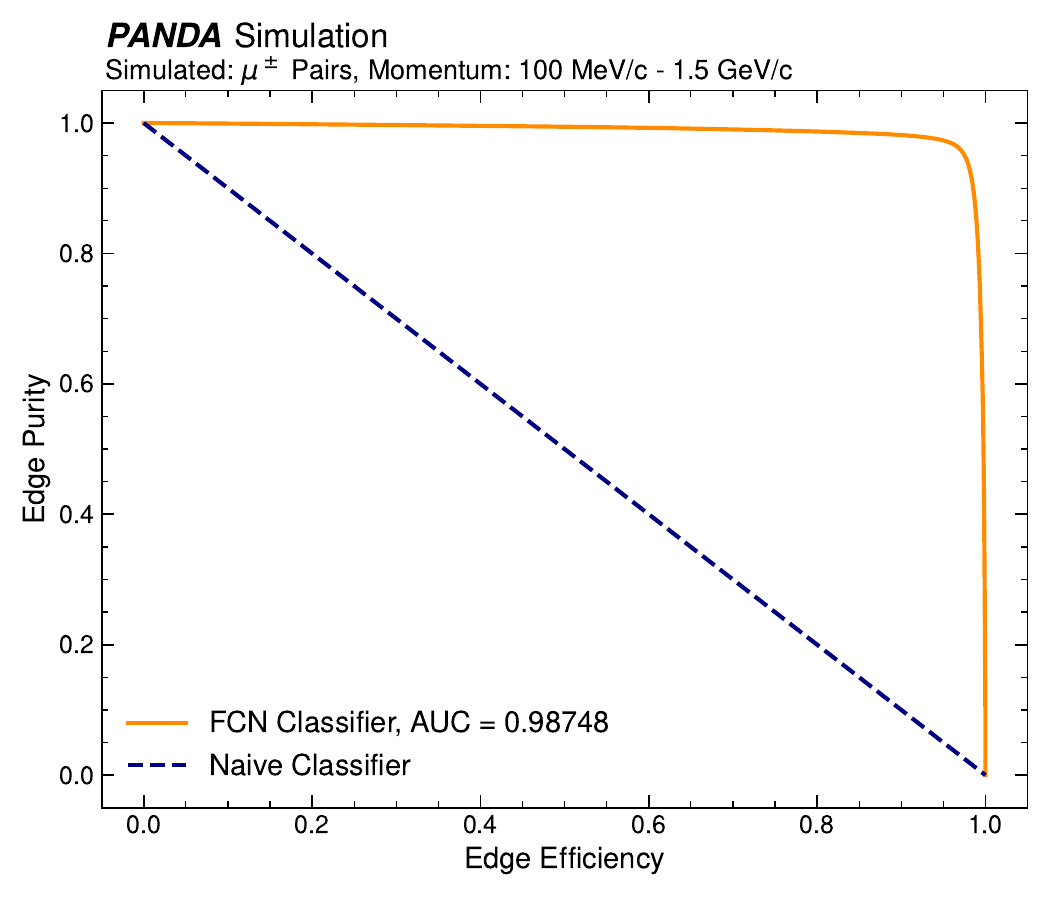}}
    \hfill
    \subfloat{\includegraphics[width=0.49\linewidth]{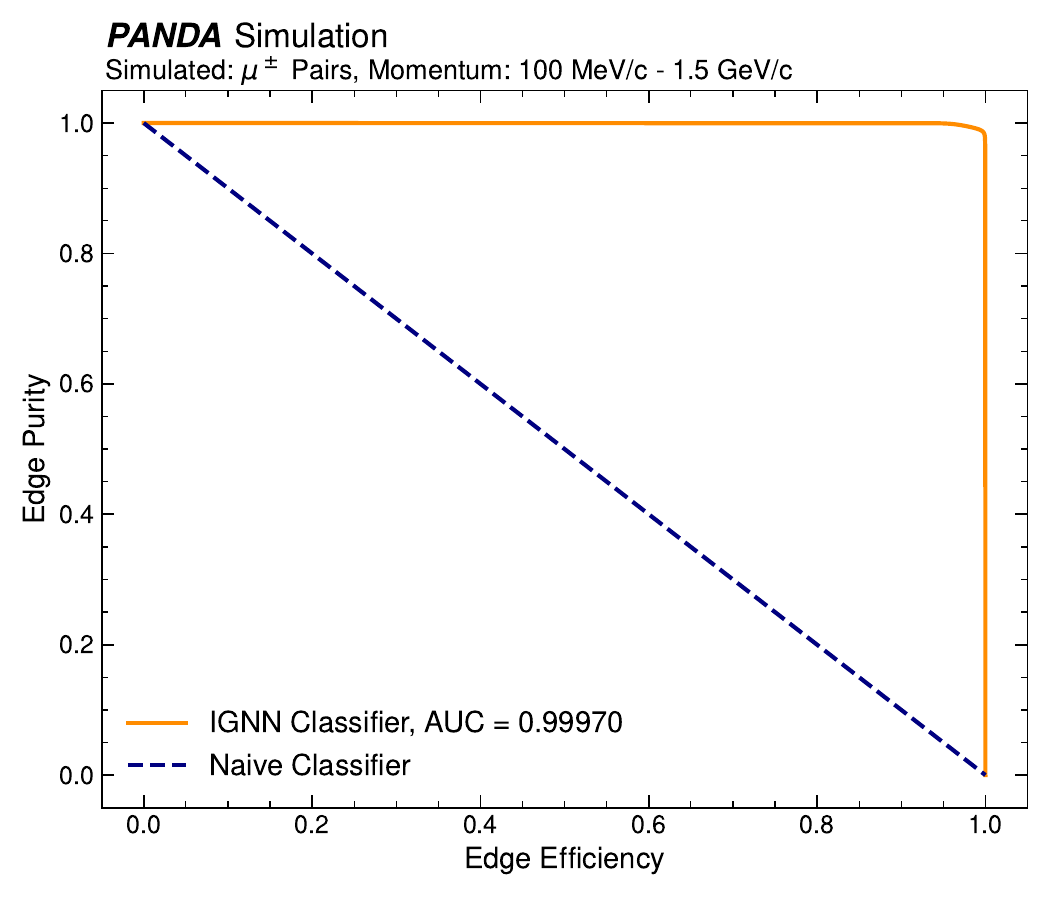}}
    \caption{ROC curve with AUC of 0.98748 for FCN (left) and 0.99970 for IGNN (right).}
    \label{fig:muon_epc_auc}
\end{figure}

\noindent IGNN gives better separation power between true and false edges compared to the FCN for a particular threshold value. To quantitatively evaluate model performance, we used the Receiving Operating Characteristic Curve (ROC) and measured Area Under the Curve (AUC). The ROC curve is constructed using the edge classification efficiency ($\epsilon_E \equiv \text{TPR}$) and edge classification purity ($\rho_E \equiv 1 - \text{FPR}$) for various thresholds. The ROC curves along with the AUCs for both models are shown in Fig. \ref{fig:muon_epc_auc}.

\begin{figure}[!ht]
    \centering
        \centering
    \subfloat{\includegraphics[width=0.5\linewidth]{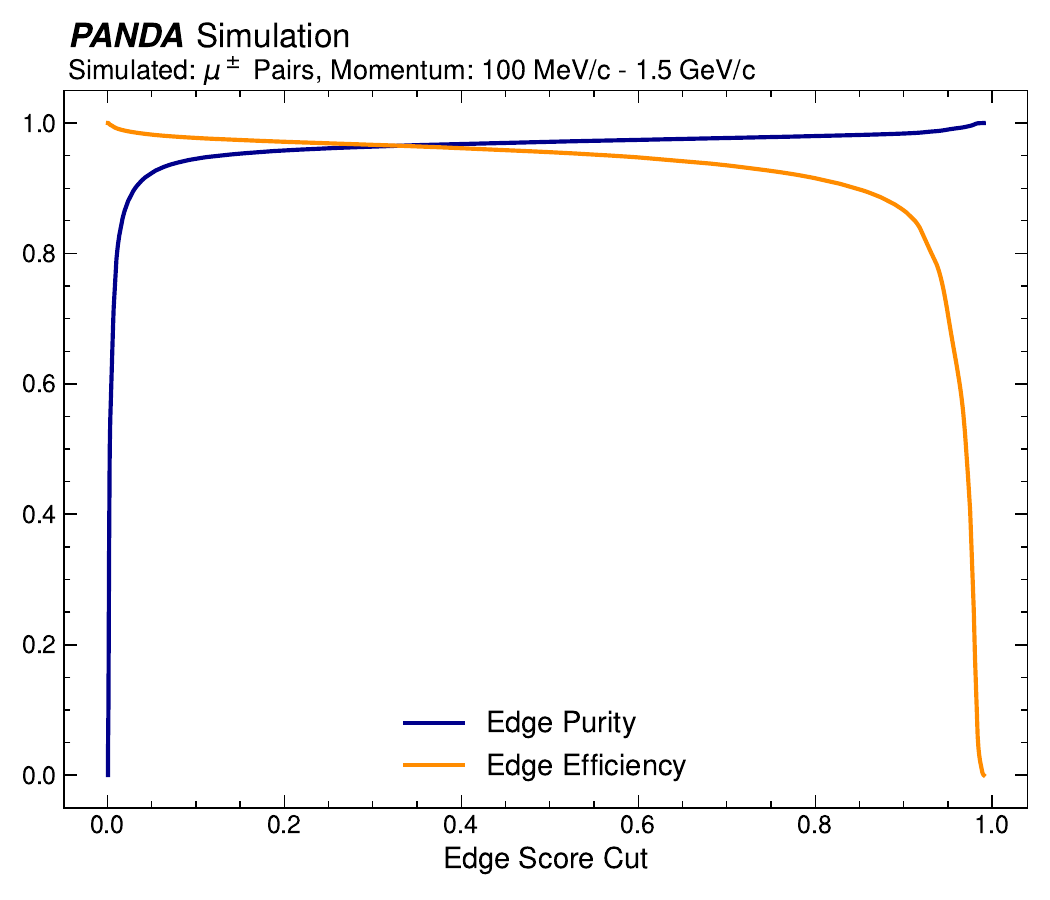}}
    \hfill
    \subfloat{\includegraphics[width=0.5\linewidth]{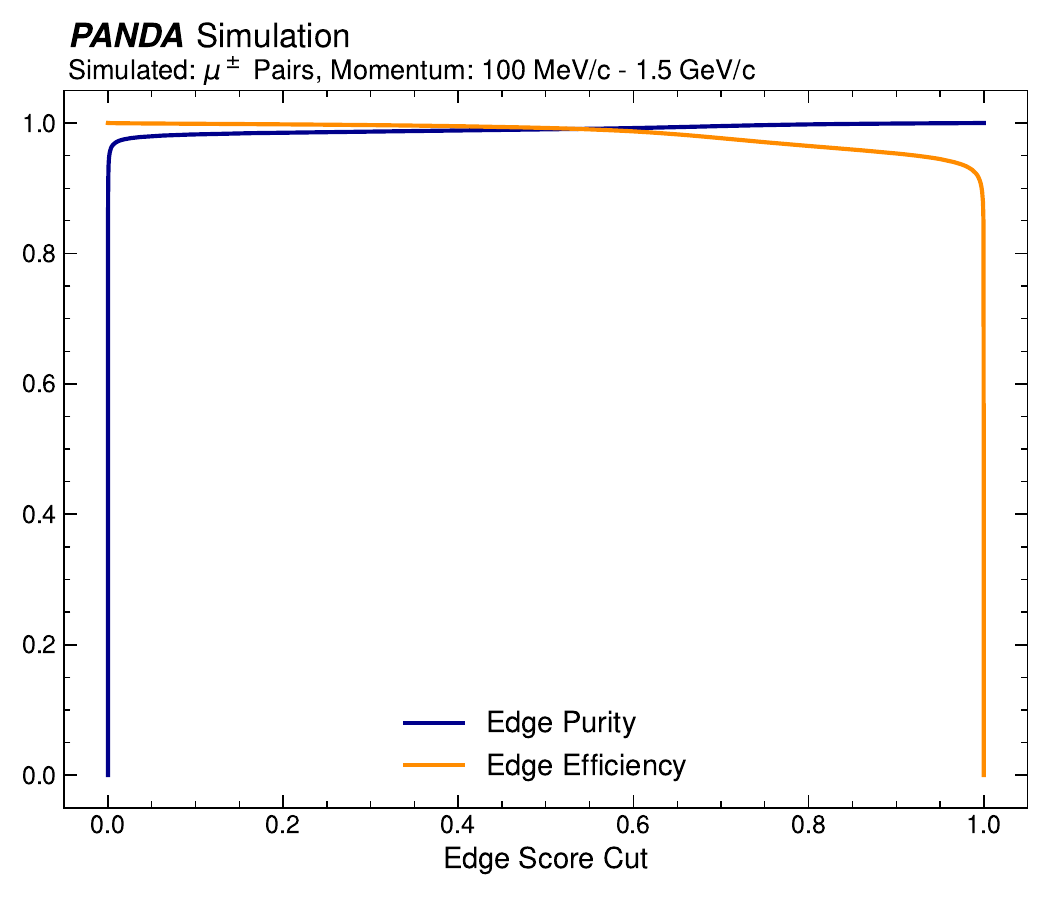}}    
    \caption{The $\epsilon_E$ and $\rho_E$ as a function of $s$ for FCN (left) and IGNN (right).}
    \label{fig:muon_epc_cut}
\end{figure}

\noindent A high value of AUC represents high model performance and vice versa, thus prompting a reasonable model training period. Since the ROC curve is constructed at varying threshold values on the model output, one needs to find an optimal threshold value or edge score cut ($s$). For this purpose, the $\epsilon_E$ and $\rho_E$ are plotted as a function of edge score cut as shown in Fig. \ref{fig:muon_epc_cut} for FCN (left) as well as IGNN (right).

\begin{figure}[!ht]
    \centering
    \subfloat{\includegraphics[width=0.49\linewidth]{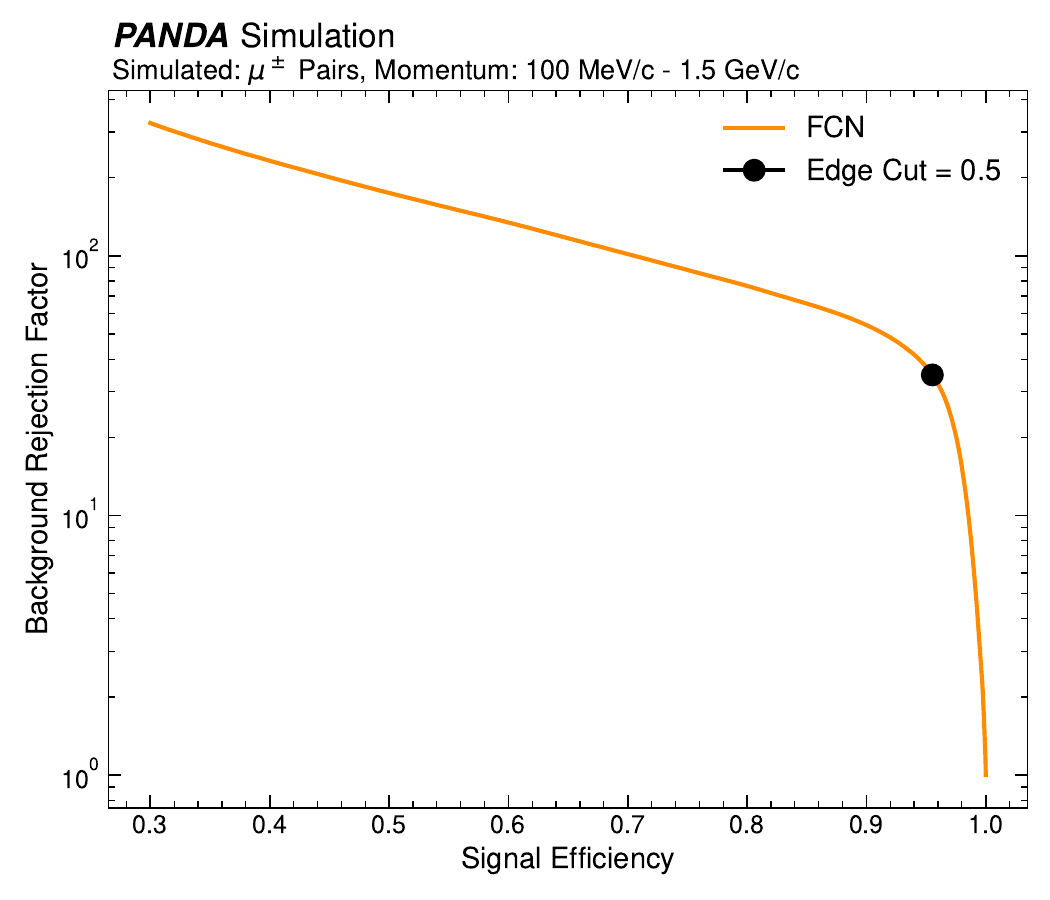}}
    \hfill
    \subfloat{\includegraphics[width=0.49\linewidth]{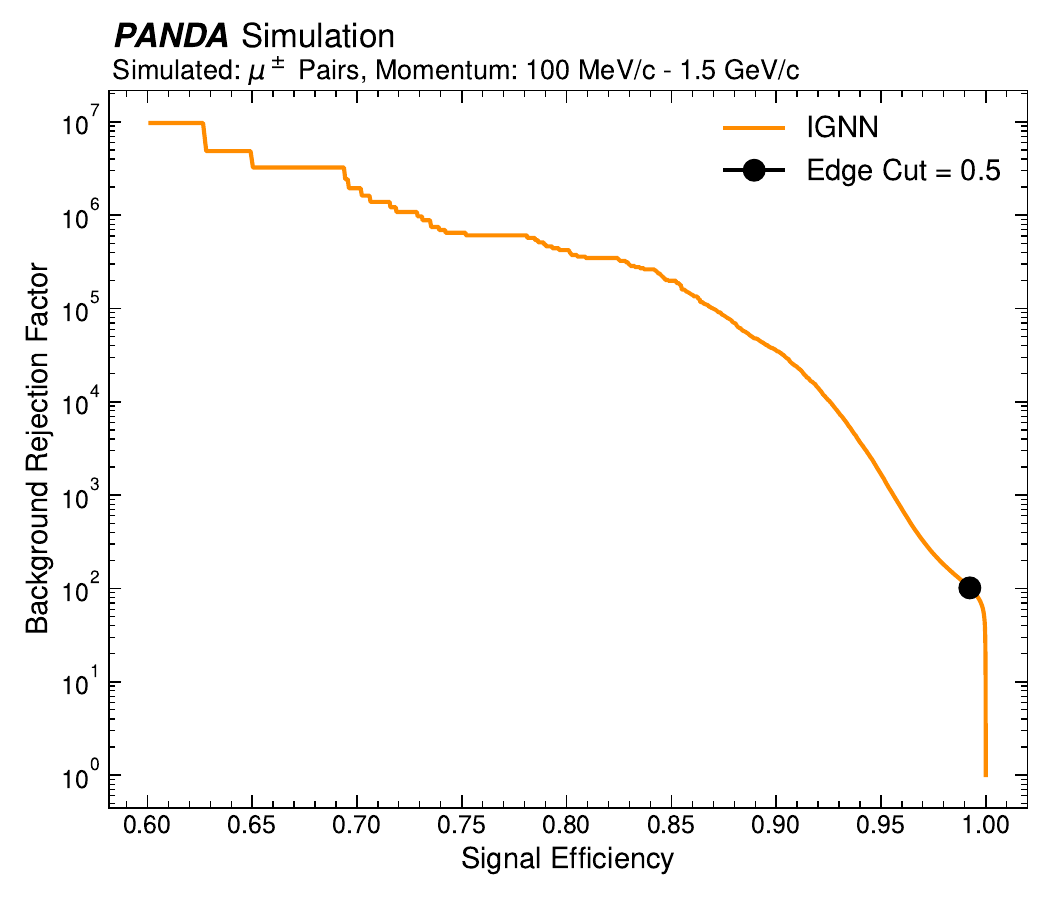}}
    \caption{The $\epsilon_{sig}$ as a function of BRF for various values of $s$ for FCN (left) and IGNN (right) models, the dot represents $s=0.5$.}
    \label{fig:muon_SB}
\end{figure}

\noindent The higher values of $s$ give high edge purity but low edge efficiency and vice versa, hence there is a trade-off in choosing a particular value of $s$. For example, choosing $s = 0.5$ gives $\epsilon_E \sim 96\%$ and $\rho_E \sim 97\%$ for the FCN model, whereas $\epsilon_E = 99.2\%$ and $\rho_E = 99.0\%$ for the IGNN. Alternatively, we can examine the signal efficiency ($\epsilon_{sig}$) vs background rejection factor (BRF) at various values of edge score cut. We define signal efficiency as the true positive rate or recall, and misidentification rate as the false positive rate from the ROC curve. The BRF is defined as the inverse of the misidentification rate. Fig. \ref{fig:muon_SB} shows the signal efficiency as a function of the BRF for various values of edge score cut for FCN (left) and IGNN (right) models.

\noindent The orange curve shows how the signal efficiency and the BRF depend on $s$ and the black dot represents the edge score cut value of $0.5$. With this cut value, we get $\epsilon_{sig} = 95.5\%$ and $\text{BRF}=34.8$ for FCN, and the $\epsilon_E$ is $99.2\%$ and the BRF of $101.4$ for IGNN. Increasing the cut value to 0.7 yields $\epsilon_{sig} = 93.5\%$ and $\text{BRF}=43.4$ for FCN and $\epsilon_{sig} = 97.7\%$ and $\text{BRF}=213.5$ for IGNN. Hence, the high BRF comes at a cost of a reduced value of $\epsilon_{sig}$. Therefore, we chose $s=0.5$ for further analysis.



\begin{figure}[!ht]
    \centering
    \subfloat{\includegraphics[width=0.49\linewidth]{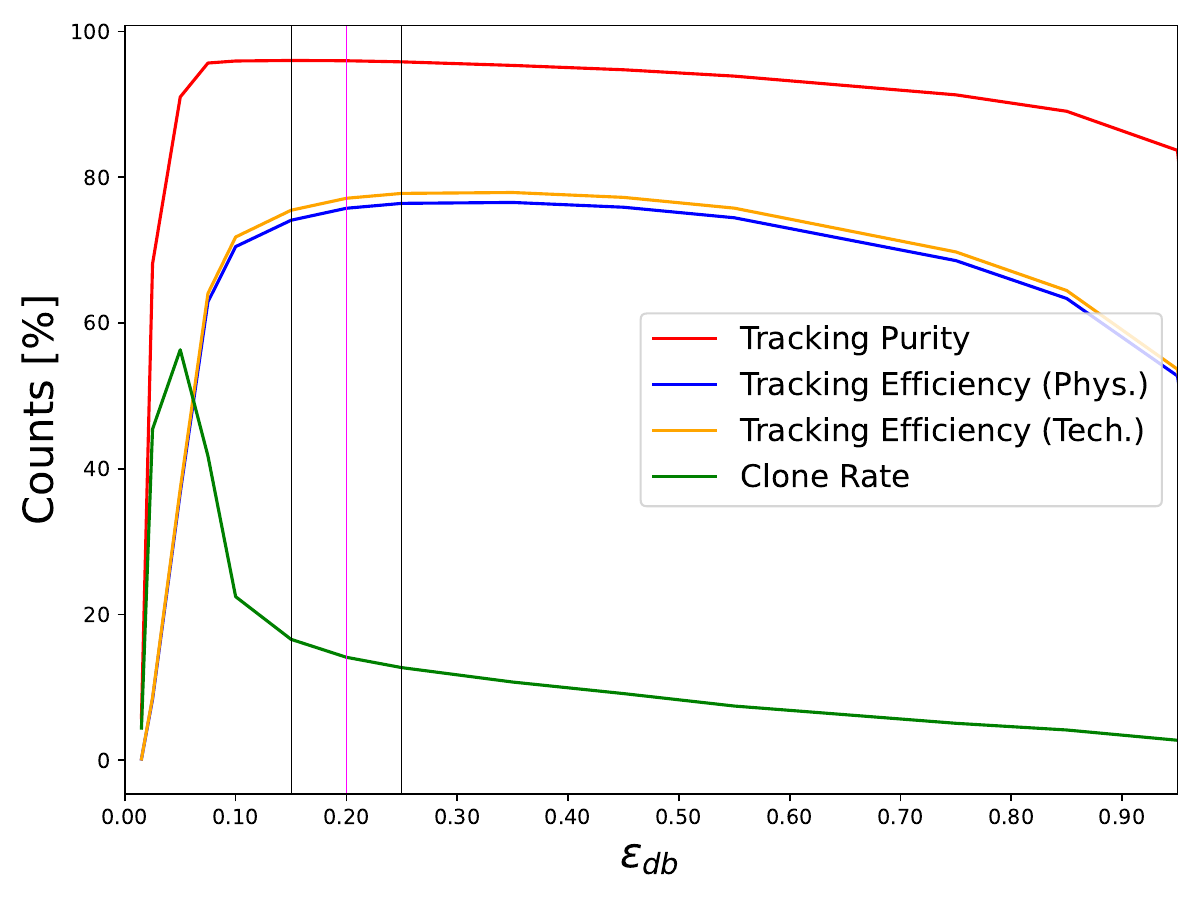}}
    \hfill
    \subfloat{\includegraphics[width=0.49\linewidth]{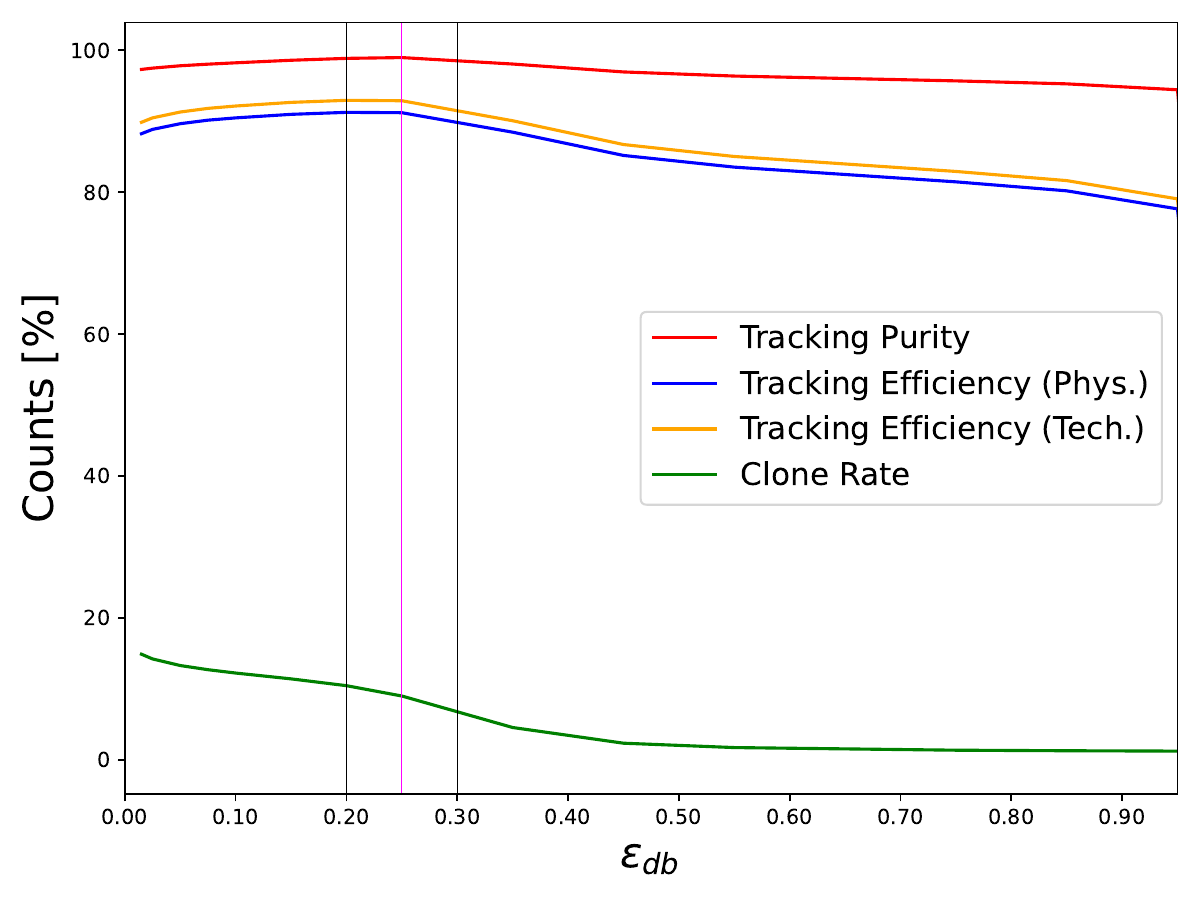}}
    \caption{The scan of $\epsilon_{db}$ for DBSCAN algorithm for FCN (left) and IGNN (right). The magenta lines show the selected values of $\epsilon_{db}$.}
    \label{fig:muon_DBSCAN}
\end{figure}

After edge labelling, we look into the graph segmentation using the DBSCAN algorithm. This algorithm requires an optimal value of the distance metric ($\epsilon_{\text{db}}$) between two nodes to cluster them together. Fig. \ref{fig:muon_DBSCAN} shows a scan of $\epsilon_{\text{db}}$ against different tracking metrics for FCN (left) and IGNN (right).

\noindent We utilize a prediction dataset comprising $2 \cdot 10^4$ events for this purpose. The optimal values of $\epsilon_{\text{db}}$ (shown as magenta vertical lines) are selected as 0.20 and 0.25 for FCN and IGNN, respectively. DBSCAN then extracts the connected components (Euclidean case) and weakly connected components (non-Euclidean case), of the graphs. The connected components are the desired track candidates. 


Finally, we evaluate the track candidates using the tracking efficiency and purity metrics discussed in Section \ref{sec:track_eval}. The tracking metrics are given in \autoref{tab:muon_track_eval} using track evaluation criteria of $N_t \ge 7$, $N_r \ge 5$ and MF $> 50\%$. 

\begin{table}[!ht]
\caption{The tracking efficiencies, ghost rate (GR), and clone rate (CR) using track evaluation criteria of $N_t \ge 7$, $N_r \ge 5$ and $\text{MF} > 50\%$.}
\label{tab:muon_track_eval}
\begin{tabular}{@{}ccccc@{}}
    \toprule[0,5pt]
    Pipeline & $\epsilon_{phys.}$ [\%] & $\epsilon_{tech.}$ [\%] & Ghost Rate [\%] & Clone Rate [\%]  \\ 
    \midrule[0,5pt]
    Deep Learning & $76.3 \pm 0.3$ & $77.2 \pm 0.3$ & $3.6 \pm 0.3$ & $17.2 \pm 0.1$  \\
    Geometric Deep Learning & $91.0 \pm 0.3$ & $92.6 \pm 0.3$ & $1.3 \pm 0.3$ & $11.5 \pm 0.1$  \\
    \bottomrule[0,5pt]
\end{tabular}
\end{table}

\noindent In the FCN case, we note that the efficiencies are fairly small from a physics perspective; for example, for events with four tracks, a tracking efficiency of $\approx$ 77\% will result in a three-fold reduction of the total efficiency. Hence, there is room for improvement. One major issue for FCN is to handle the huge class imbalance with a ratio of true to false edges of 1:4. IGNN can handle such imbalances by aggregating neighbourhood relations through message-passing, which results in an increase from 77.2\% to 92.6\% in the tracking efficiencies, an almost $20\%$ increase in efficiencies. Furthermore, it reduces the ghost rate to be almost negligible. The clone rate is also reduced, but still high.

\begin{figure}[!htb]
    \centering
    \subfloat{\includegraphics[width=0.49\linewidth]{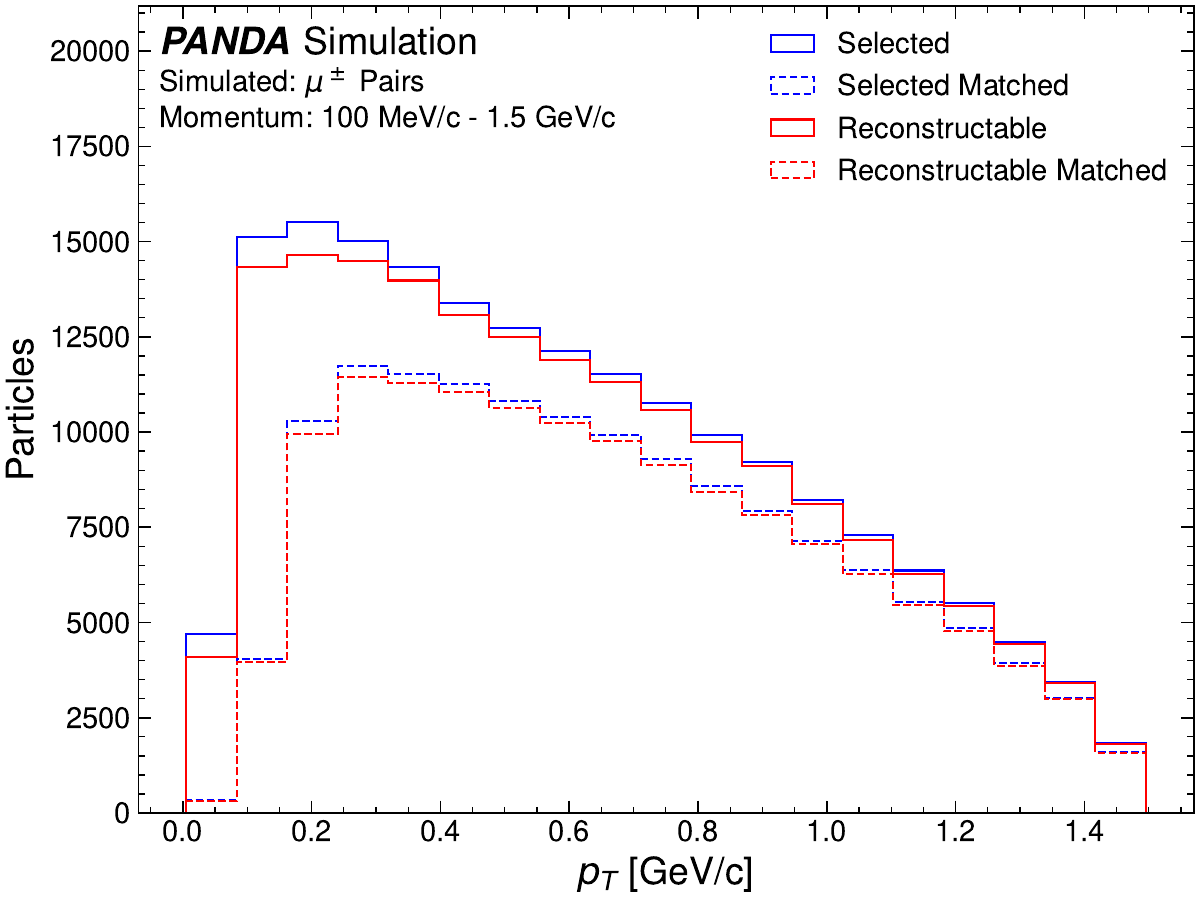}}
    \hfill
    \subfloat{\includegraphics[width=0.49\linewidth]{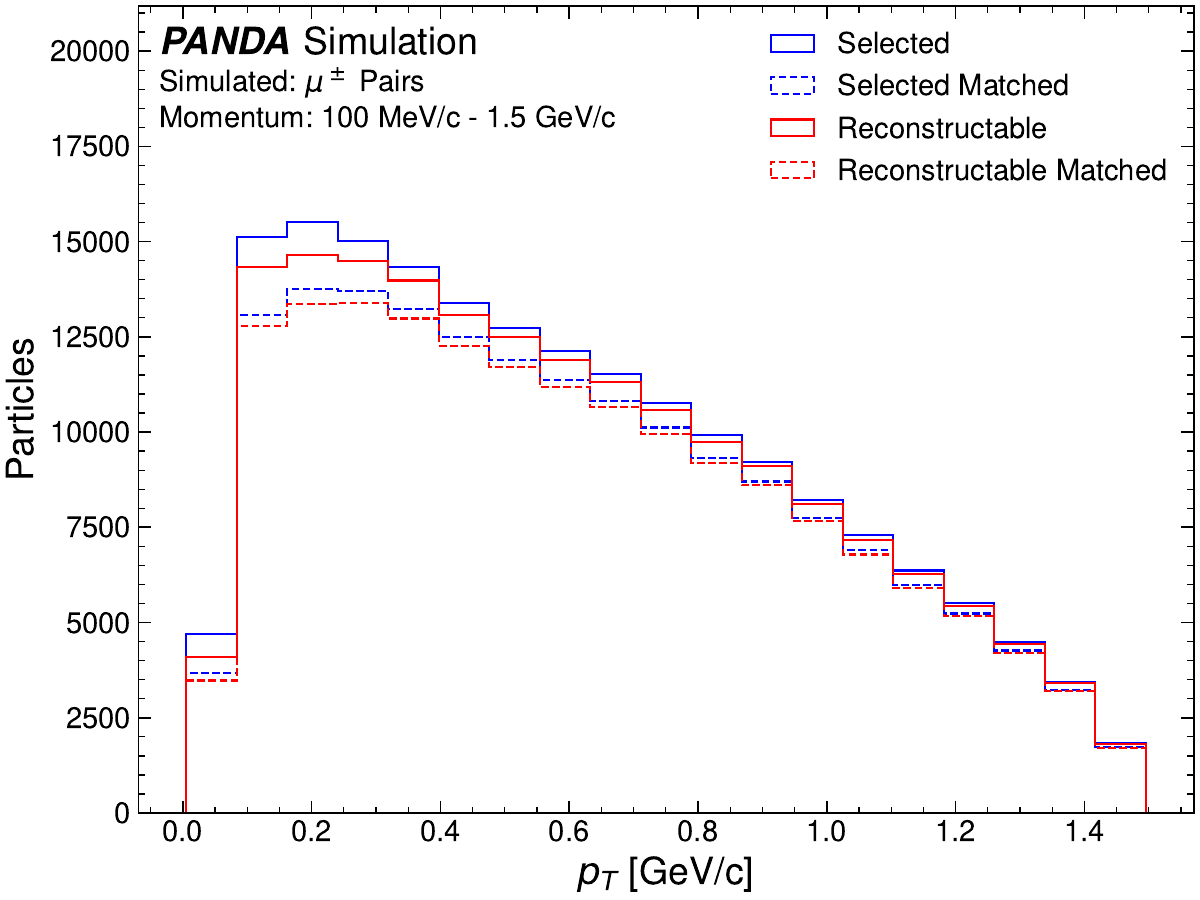}}
     \caption{The number of selected, selected and matched, reconstructable, reconstructable and matched particles as a function of \pt for FCN (left) and IGNN (right).}
    \label{fig:muon_particles_pt}
\end{figure}

\begin{figure}[!htb]
    \centering
    \subfloat{\includegraphics[width=0.49\linewidth]{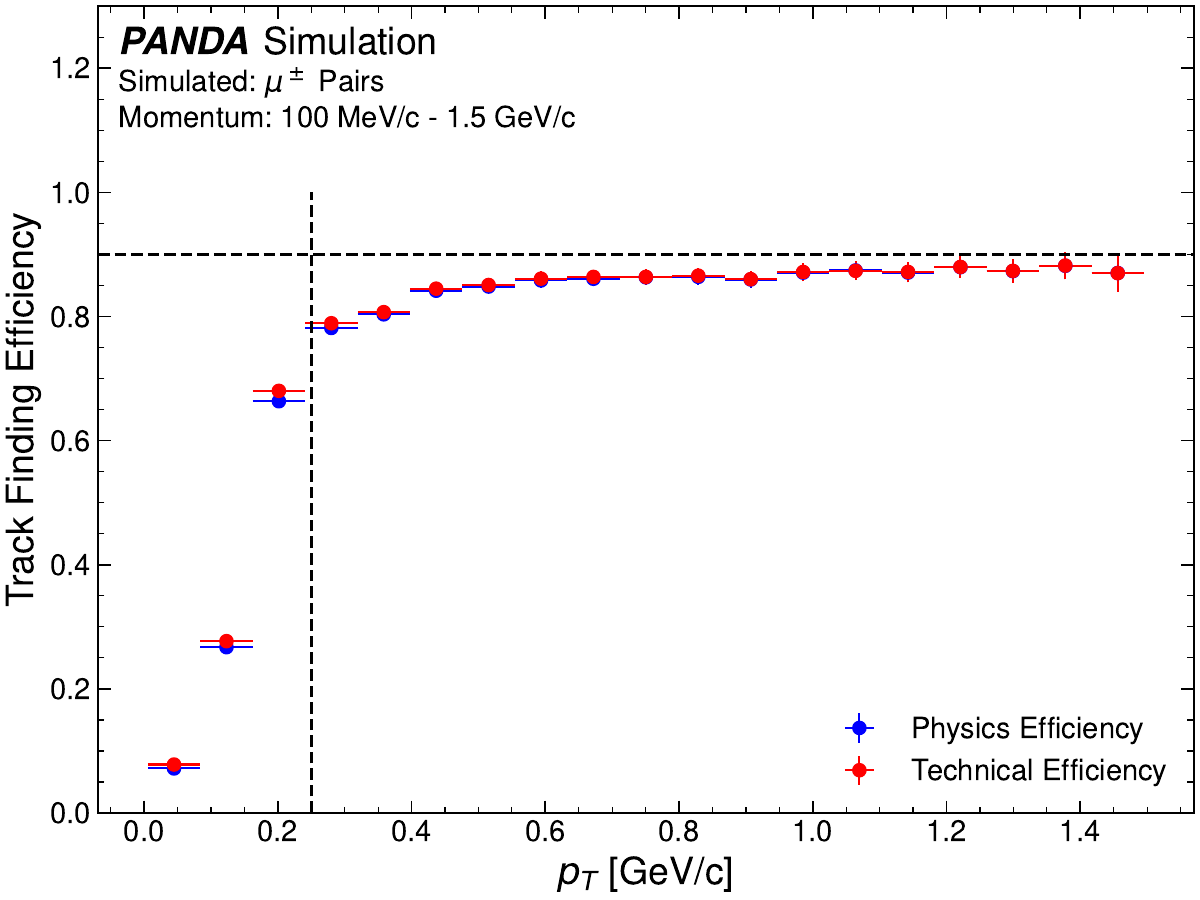}}
    \hfill
    \subfloat{\includegraphics[width=0.49\linewidth]{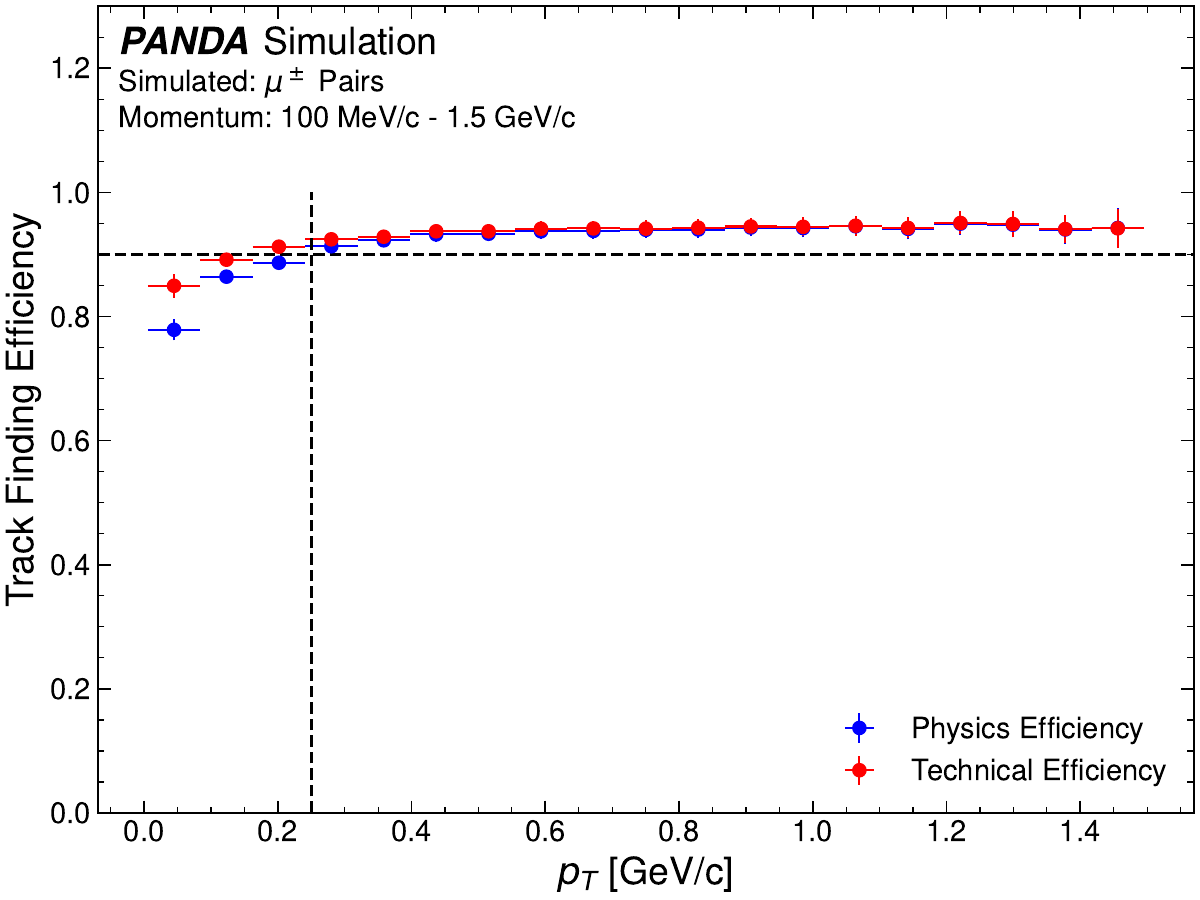}}
    \caption{Tracking efficiencies as a function of \pt for FCN (left) and IGNN (right) with reference lines at $\pt = 0.25$ GeV/c (vertical) and $\epsilon = 90\%$ efficiency (horizontal).}
    \label{fig:muon_efficiencies_pt}
\end{figure}

To better understand the algorithm's performance, we investigate how the tracking efficiency depends on the transverse momentum (\pt) of muons.  Fig. \ref{fig:muon_particles_pt} shows the number of particles (selected, selected and matched, reconstructable and reconstructable and matched) for FCN (left panel) and IGNN (right panel) as a function of \pt. In Fig. \ref{fig:muon_efficiencies_pt}, we show the corresponding track efficiencies. We conclude that the main loss of tracks occurs at low \pt, especially below $\pt = 0.25$ GeV/c. Particles with such low \pt have trajectories with large enough curvature that they make a turn before traversing the full STT detector. Hence, they are trapped inside the detector, with trajectories spiralling in the magnetic field, lose energy in interactions with the detector material and the gas, and potentially also intersect the trajectories of other particles. This is in contrast to high \pt particles, which have rather straight trajectories with fewer intersections resulting in higher efficiencies for this class of particles. The improved performance of the IGNN compared to the FCN for low-$\pt$ tracks is striking, as seen in Fig. \ref{fig:muon_efficiencies_pt}. 

\subsection{Hyperon Reconstruction with GDL}\label{sec:fwp_gdl}

\begin{figure}[!htb]
    \centering
    \subfloat{\includegraphics[width=0.49\linewidth]{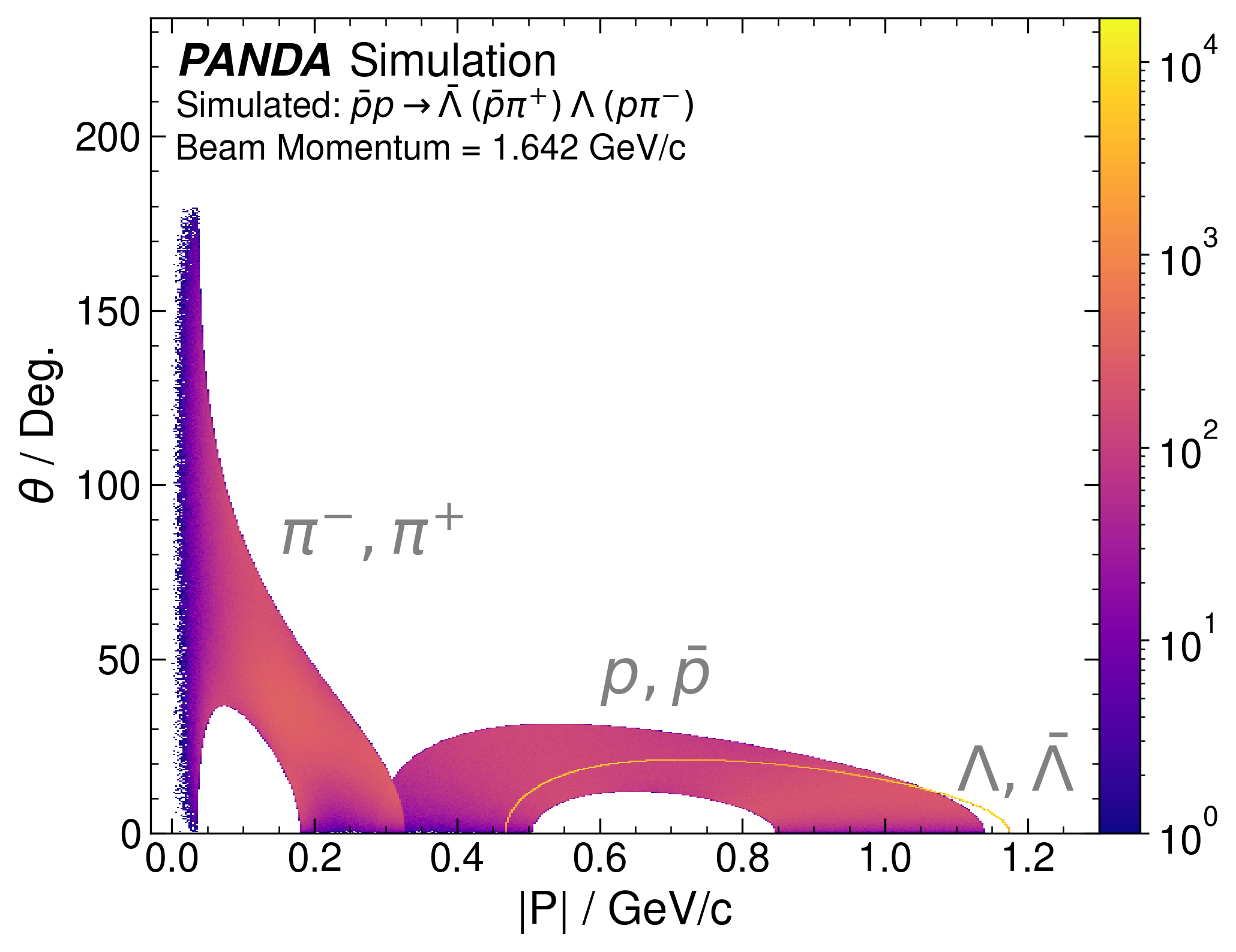}}
    \hfill
    \subfloat{\includegraphics[width=0.49\linewidth]{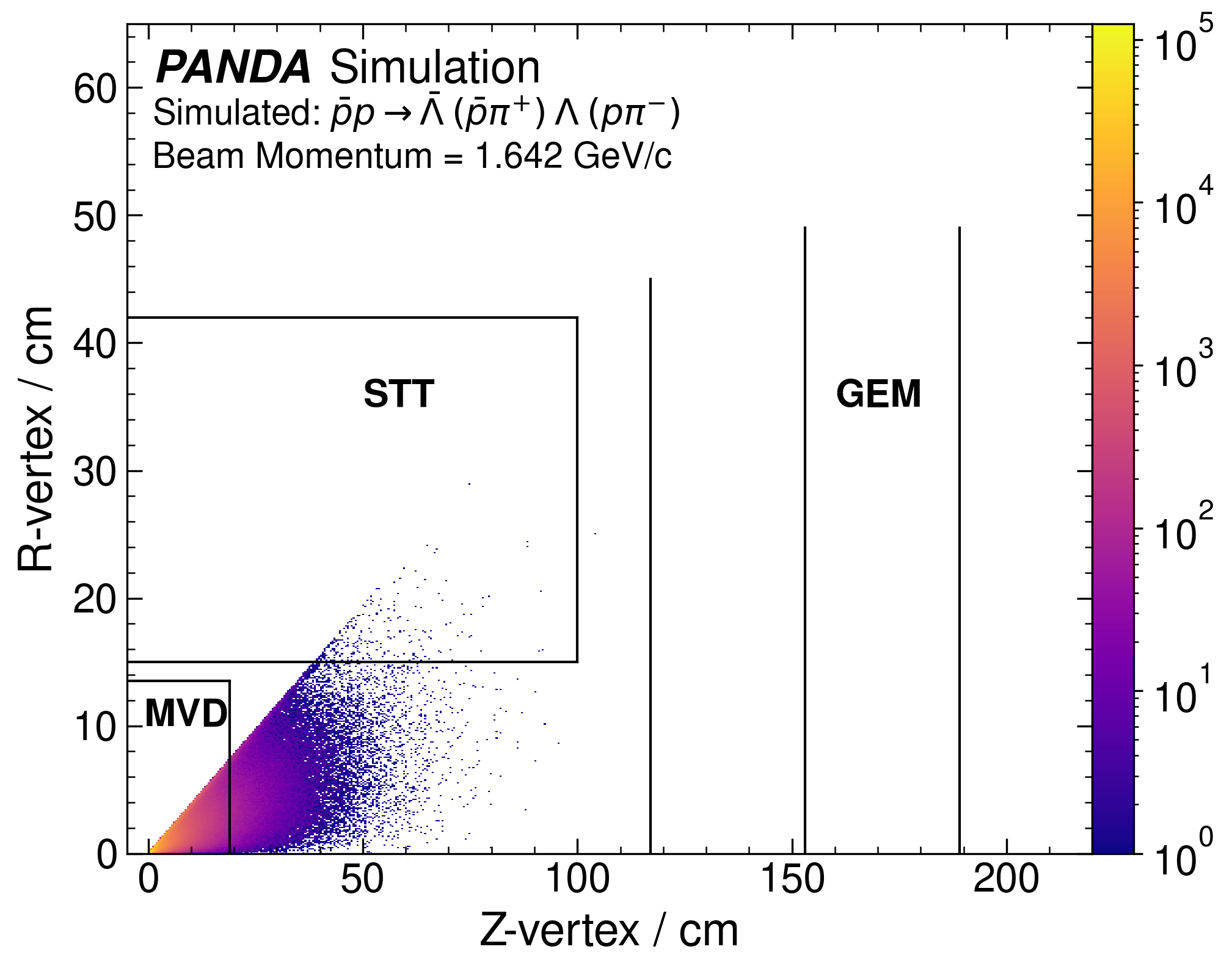}}
    \caption{Kinematic properties of $\bar{p}p \rightarrow \bar{\Lambda}\Lambda \rightarrow \bar{p}\pi^{+}p\pi^{-}$ reaction: (left) MC $|p|$ \textit{versus} $\theta$ distribution, (right) MC decay vertex distribution of $\Lambda$ and $\bar{\Lambda}$.}
    \label{fig:sim}
\end{figure}

We have performed simulations for training and testing with the reaction $\bar{p}p \rightarrow \bar{\Lambda}\Lambda \rightarrow \bar{p}\pi^{+}p\pi^{-}$ at a beam momentum of $\bar{p}_{beam} = 1.64$ GeV/$c$. Since $\Lambda$ hyperons are neutral, tracking information is obtained from their charged daughters, \textit{i.e.} protons and pions. The $\bar{p}p \rightarrow \bar{\Lambda}\Lambda$ reaction has been rigorously studied by the PS185 experiment at LEAR, CERN \cite{Johansson:1999emx, Barnes:2000be}, in particular at this beam momentum \cite{PS185:2006yyx}. It has been found that in the CMS system of the reaction, the $\bar{\Lambda}$ antihyperon is emitted predominantly in the forward direction while the $\Lambda$ hyperons go backwards. This means that in the lab system of PANDA, the fast $\bar{\Lambda}$ antihyperon goes into the acceptance of the Forward Spectrometer while the $\Lambda$ hyperons are slow and decay inside the Target Spectrometer. Hence, the daughters of the $\Lambda$ give rise to hits in the STT and can be reconstructed with our algorithm. Of special interest is the daughter pions from $\Lambda$ decays, since they often have very low momenta (see left panel of Fig. \ref{fig:sim}): In the decay, antiproton and protons ($\bar{p}, p$) take the larger share of the momentum, while only a small fraction goes to the pions ($\pi^+, \pi^-$). These pions are challenging to reconstruct due to the high curvature of their trajectories and their high probability of intersecting with the trajectories of other particles. Furthermore, due to the relatively long lifetime of the $\Lambda$ hyperons, they are expected to decay far from the beam-target interaction point (see right panel of Fig. \ref{fig:sim}). This makes the $\bar{p}p \rightarrow \bar{\Lambda}\Lambda \rightarrow \bar{p}\pi^{+}p\pi^{-}$ reaction an important benchmark for track reconstruction algorithms with PANDA.

\noindent For the hyperon reconstruction, we applied the same GDL pipeline as in the muon case except with one small difference: in the graph construction stage, the heuristic method for building nodes and edges was not restricted to adjacent sectors. Since the $\bar{p}p \rightarrow \bar{\Lambda}\Lambda \rightarrow \bar{p}\pi^{+}p\pi^{-}$ reaction contains fewer particles per event compared to the $5\mu^+\mu^-$ case, and since we expect many pions to be emitted at extremely low \pt \cite{PANDA:2023ljx}, removing this condition increases the amount of data in each event. After graph labelling, we tested $2 \cdot 10^3$ events during inference. In the graph segmentation stage, we used the DBSCAN method with $\epsilon_{\text{db}} = 0.15$, and a minimum number of samples to be two to find the connected components from the test events. Using the same track evaluation criteria as in previous cases, the tracking efficiencies, ghost rate and clone rate are obtained as in \autoref{tab:fwp_track_eval}.

\begin{table}[!ht]
\caption{Tracking efficiencies, ghost rate, and clone rate for hyperons.}\label{tab:fwp_track_eval}
\begin{tabular}{@{}ccccccc@{}}
    \toprule[0,5pt]
    $N_t$ & $N_r$ & MF [\%] & $\epsilon_{phys.}$ [\%] & $\epsilon_{tech.}$ [\%] & Ghost Rate [\%] & Clone Rate [\%]  \\ \midrule[0,5pt]
    7 & 5 & $> 50$ & $89.6 \pm 0.6$ & $97.1 \pm 0.6$ & $0.5 \pm 0.6$ & $4.9 \pm 0.1$  \\ \bottomrule[0,5pt]
\end{tabular}
\end{table}

\noindent The physics tracking efficiency $\epsilon_{phys.}$ is about 90\%. However, the technical efficiency $\epsilon_{tech.}$ is significantly higher and the low ghost rate and clone rate are significantly lower than in the multi-muon case. This performance gain is understood as each event has fewer particles than the muon case and there are fewer track intersections. 

\begin{figure}[!htb]
    \centering
    \subfloat{\includegraphics[width=0.5\linewidth]{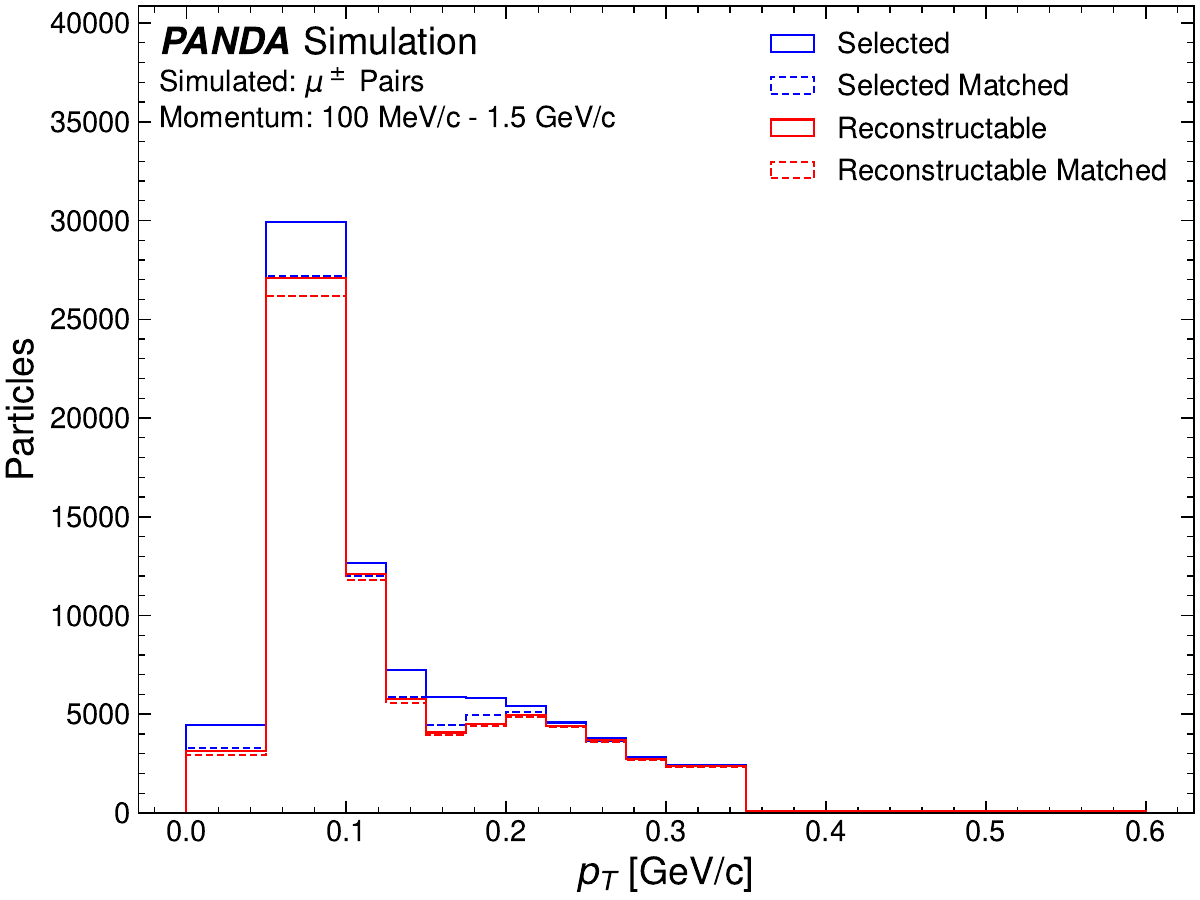}}
    \hfill
    \subfloat{\includegraphics[width=0.5\linewidth]{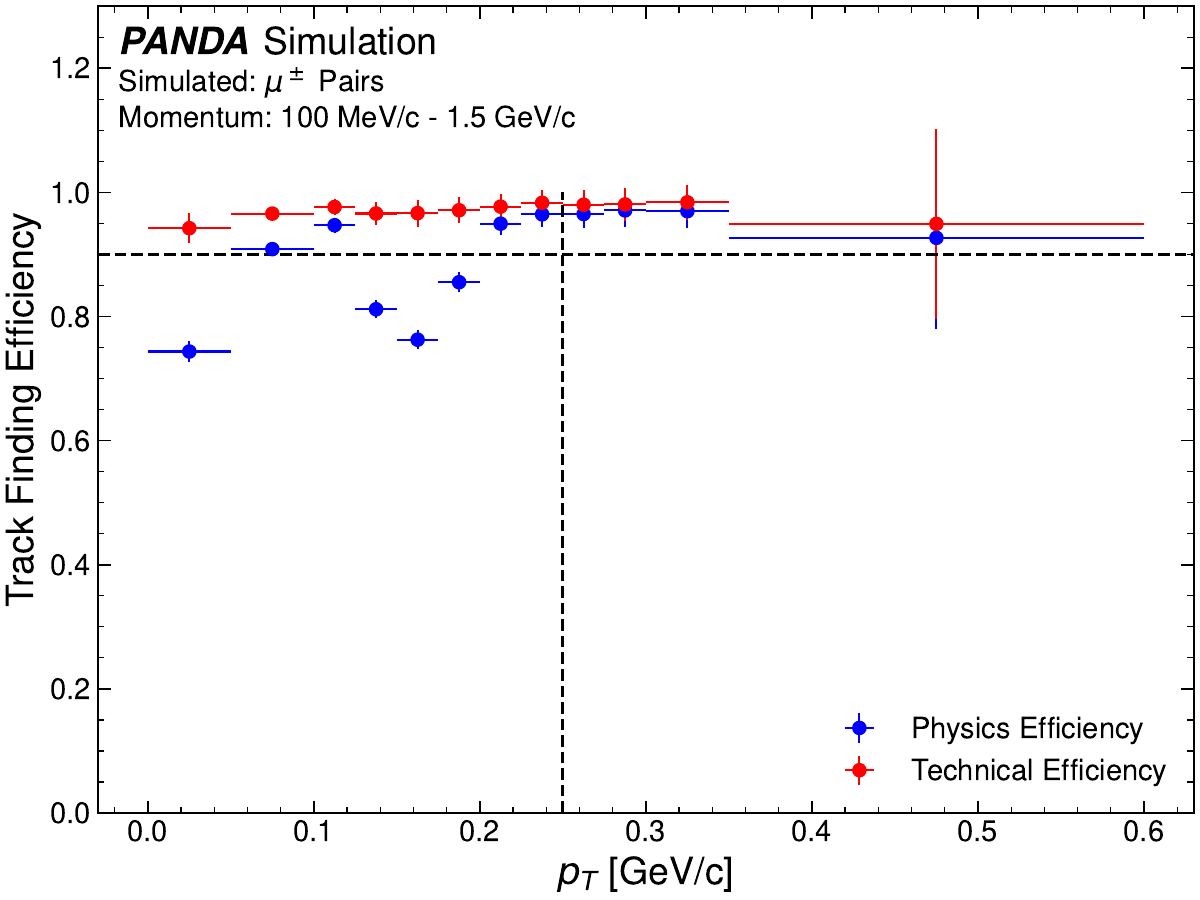}}
    \caption{Number of particles (left), and tracking efficiencies (right) as a function of \pt with reference lines at $\pt = 0.25$ GeV/c (vertical) and $\epsilon = 90\%$ efficiency (horizontal).}
    \label{fig:fwp_pt_plots}
\end{figure}

Similar to the muon case, we analyse the tracking efficiencies as a function of \pt as shown in Fig. \ref{fig:fwp_pt_plots}: the number of particles (left) and tracking efficiencies (right).

\noindent In a large fraction of the events, there are particles momenta as low as \pt $< 0.25$ GeV/c which are captured in the magnetic field of the PANDA solenoid and therefore remain inside the detector. These particles are primarily pions and form an enhancement at low \pt in the left panel of Fig. \ref{fig:fwp_pt_plots}. Protons generally have much larger momenta, and the different kinematics of protons and pions manifest in different lab polar angles. As a result, the track lengths will differ and thus the reconstruction probability. In the right panel of Fig. \ref{fig:fwp_pt_plots}, we see that the physical track efficiency $\epsilon_{phys}$ has a structure where low-momentum pions and high-momentum protons are relatively well reconstructed. At the same time, there is a dip in the efficiency in the intermediate region.  However, the technical efficiency has no such structure which leads to the conclusion that the intermediate \pt region has a high content of non-reconstructable tracks.

\begin{figure}[!htb]
    \centering
    \subfloat{\includegraphics[width=0.5\linewidth]{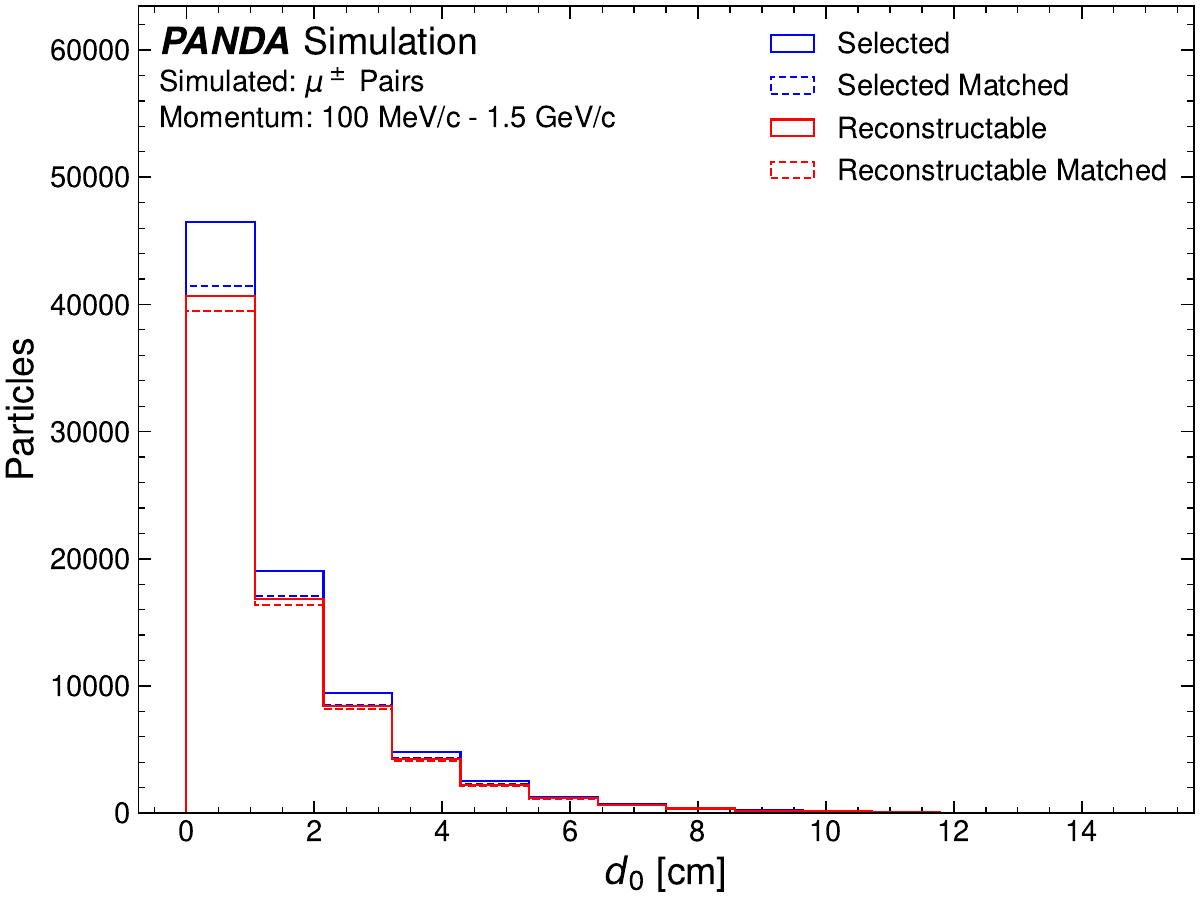}}
    \hfill
    \subfloat{\includegraphics[width=0.5\linewidth]{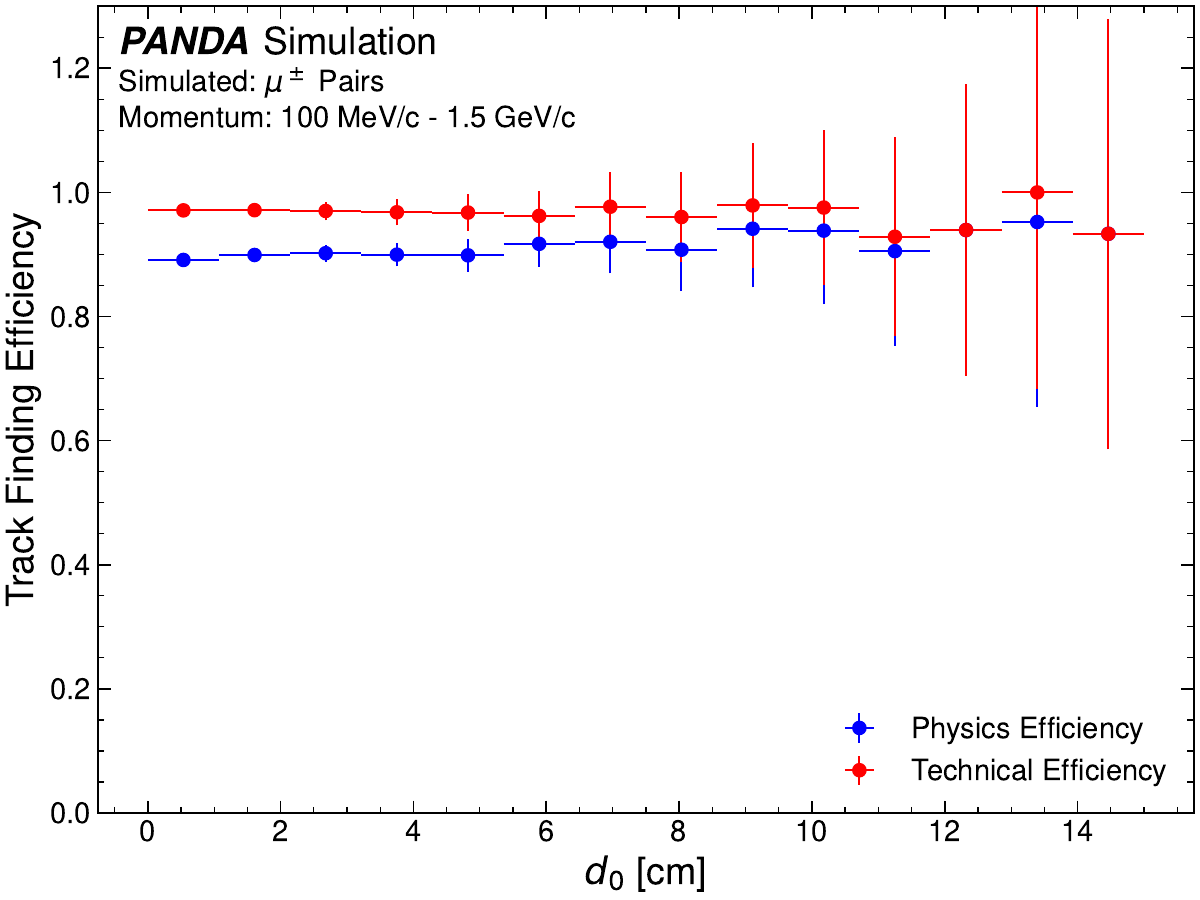}}
    \caption{Number of particles (left), and tracking efficiencies (right) as a function of $\text{d}_0$.}
    \label{fig:fwp_d0_plots}
\end{figure}

Next, we investigate tracking efficiency as a function of the radial position of decay vertices ($\text{d}_0$), as shown in Fig. \ref{fig:fwp_d0_plots}. 

\noindent Most particles (protons, pions) are generated close to the interaction point, however, a considerable fraction is generated up to 14 cm from the beam-target interaction point. From Fig. \ref{fig:fwp_d0_plots}, we conclude that our algorithm also performs well for these kinds of tracks: the physical and technical track efficiencies are above 90\% for both pions and protons over the full $\text{d}_0$ range. Moreover, the technical efficiency is about 97\%. This is an important finding as most heavier hyperons ($\Xi, \Omega$, etc.) decay at $\text{d}_0 < 15$ cm \cite{PANDA:2023ljx}.



\section{Conclusion}\label{sec:conclusion}

In this work, we have successfully applied machine learning to reconstruct particle trajectories in a hadron physics experiment. Our work shows the first use of GNNs-based track reconstruction in the straw tube detector with non-Euclidean geometry. It is found that GDL models give promising results giving overall tracking efficiency $\ge 90\%$. It can reconstruct pions with \pt as low as $\sim 0.05$ GeV/c and protons with \pt as low as $\sim 0.1$ GeV/c. Further studies show that this method also works well for reconstructing particles with secondary decay vertices up to at least $\text{d}_0 = 14$ cm away from the IP in the radial direction. Beyond $\text{d}_0 = 14$ cm, our simulated data contains no decaying $\Lambda$ hyperons. This is an important result as heavier hyperons, such as $\Xi^-$ and $\Omega^-$, are expected to decay through intermediate $\Lambda$ hyperons with the $\Lambda$ decay vertices mostly occurring less than $15$ cm. These results are promising for the hyperon reconstruction at PANDA, and demonstrate the virtues of GNNs for the specific challenges of particle tracking in hadron physics experiments.


\section{Supplementary information}

The relevant code used for this work is available at \cite{adeel_akram_2025_15024201}.

\section*{Conflict of Interest}
The authors declare that we have no conflicts of interest.

\section*{Acknowledgments}
The authors would like to thank Nikolai in der Wiesche for productive discussions. This project has received funding from the Knut and Alice Wallenberg Foundation and the Swedish Research Council (Sweden).









\bibliography{sn-bibliography}

\end{document}